\documentclass[letterpaper]{JHEP3}
\usepackage{nicefrac}

\usepackage{amsmath,epsfig}
\usepackage{amssymb,amsfonts}

\usepackage{graphics}
\usepackage{epsfig}
\usepackage{amsfonts}

\renewcommand{\Im}{\mbox{Im}}
\renewcommand{\Re}{\mbox{Re}}

\newcommand{\Tr}{\mbox{Tr}}

\def\d{\partial}

\def\s{\sigma}

\def\a{\alpha}

\def\e{\epsilon}

\def\h{\eta}

\def\half{{\frac12}}

\def\IC{\relax\hbox{$\inbar\kern-.3em{\rm C}$}}

\def\IC{{\bf C}}

\def\bea{\begin{eqnarray}}
\def\eea{\end{eqnarray}}
\def\be{\begin{equation}}
\def\ee{\end{equation}}
\def\ba{\begin{align}}
\def\ea{\end{align}}
\def\bse{\begin{subequations}}
\def\ese{\end{subequations}}
\def\1F1{{}_1\!F_1}
\def\2F0{{}_2\!F_0}

\def\a{\alpha}

\def\h3{$\textrm{H}_3^+$}

\def\d{{\partial}}

\def\IC{{\mathbb C}}

\def\Tr{{\rm Tr}}

\def\lbldef#1#2{\expandafter\gdef\csname #1\endcsname {#2}}

\def\href#1#2{#2}

\newcommand{\beq}{\begin{equation}}
\newcommand{\eeq}{\end{equation}}
\newcommand{\ber}{\begin{eqnarray}}
\newcommand{\eer}{\end{eqnarray}}

\def\be{\begin{eqnarray}}
\def\ee{\end{eqnarray}}

\def\({\left(}
\def\){\right)}
\def\[{\left[}
\def\]{\right]}
\def\<{\langle}
\def\>{\rangle}
\def\d{\partial}

\preprint{Brown-HET-1590 \\ YITP-SB-09-41}

\title{A Spin Chain for the 
 Symmetric Product CFT$_2$}

\author{
Ari Pakman\footnote{Email: ari$\_$pakman@brown.edu}$^{~1}$, Leonardo Rastelli\footnote{Email: leonardo.rastelli@stonybrook.edu}$^{~2}$, and
Shlomo S. Razamat\footnote{Email: razamat@max2.physics.sunysb.edu}$^{~2}$
\\ \\ \\
\it $^1$ Department of Physics,\\ Brown University,\\
Providence, RI 02912, USA
\\
\\
\it $^2$ C.N. Yang Institute for Theoretical Physics,\\
\it Stony Brook University, \\
\it Stony Brook, NY 11794-3840, USA}

\abstract{

We  consider ``gauge invariant''  operators in ${\rm Sym}^N \, T^4$,
 the  symmetric product  orbifold
of $N$  copies of the $2d$ supersymmetric sigma model with $T^4$ target.
We discuss a spin chain  representation for single-cycle operators and
 study their two point functions at large $N$.
We perform systematic calculations at the orbifold point (``tree level''),
where non-trivial mixing is already present, and some sample calculations 
 to first  order in the blow-up mode of the orbifold (``one loop'').
}

\keywords{CFT, AdS/CFT}

\begin{document}

\section{Introduction}


In gauge/string dualities, it is desirable to 
fix the bulk-to-boundary dictionary as precisely as possible.
 In general this is a daunting task, but there are  some very (super)symmetric examples where integrability makes the problem  more tractable.
In the maximally supersymmetric $AdS_5/CFT_4$ duality, 
the discovery of integrability both in the string sigma model~\cite{Bena:2003wd} 
and in the large $N$ field theory~\cite{Minahan:2002ve}
 has triggered spectacular progress  in calculating and  matching
 the spectrum of states. See for instance~ \cite{Beisert:2004ry,Plefka:2005bk,Minahan:2006sk,Dorey:2009zz,Arutyunov:2009ga} for reviews
 of this very vast literature.
More recently this program  has been extended to the $AdS_4/CFT_3$ duality \cite{Aharony:2008ug,Aharony:2008gk},
see {\it e.g.}~\cite{Minahan:2008hf,Gaiotto:2008cg,Bak:2008cp,Gromov:2008qe}. 
Curiously another classic duality, the $AdS_3/CFT_2$ correspondence~\cite{Maldacena:1997re}, 
has   been largely left out from these  developments. 
While on the string theory side there is evidence for integrability \cite{Chen:2005uj,Adam:2007ws},
much less is known on the field theory side. This predicament
is the background motivation for this paper.

To describe the peculiarities of  $AdS_3/CFT_2$, 
let us  briefly review  how integrability works in the higher-dimensional cases.
In  $4d$ Yang-Mills  (and also in $3d$ Chern-Simons coupled to matter),
the  elementary gauge invariant states at large $N$ are single-trace operators, which can be viewed
 as spin chain systems with periodic boundary  conditions. The dilatation operator of ${\cal N} = 4$ SYM
 is identified with an integrable spin chain Hamiltonian.
 The spin chain Hamiltonian is 
  of nearest-neighbor type  to lowest order in planar perturbation theory, and it becomes more and more non-local
  to higher orders. In fact at higher orders the Hamiltonian is not explicitly known and the most 
   effective tool is instead the S-matrix of asymptotic magnon excitations propagating on the spin chain \cite{Staudacher:2004tk}.
   An analogous S-matrix can  be defined for the light-cone string sigma model, and when phrased in this language
   the string theory analysis and field theory analysis become largely isomorphic.
   The AdS/CFT S-matrix  is fixed by symmetries and various consistency
   requirements     and is the main input in an asymptotic Bethe ansatz \cite{Beisert:2006ez},  which in principle
   allows to calculate the dimension of all ``long'' operators for arbitrary coupling. More recently a conjecture has been formulated that
   extends the Bethe ansatz to  arbitrary finite-size operators \cite{Gromov:2009tv}.

The same ideas  and techniques should apply at least to the string side of the $AdS_3/CFT_2$ duality.
However the real interest would be to develop the string and field theory side
simultaneously and understand their dictionary.  The  $CFT_2$ side is the less understood and it is the focus
of our work. In  a certain (possibly singular) region
of its moduli space, the theory is believed to be described by 
a deformation of ${\rm Sym}^N \, T^4$, see {\it e.g.} \cite{Seiberg:1999xz,Larsen:1999uk}. 
 This is the best analogy we have for the weakly coupled region of  Yang-Mills or Chern-Simons theory.
However 
 the notion of a spin chain is less obvious than in a gauge theory, and  perturbative computations (in conformal perturbation theory,
 infinitesimally away from the orbifold point) take a  qualitatively different form.
Following the ideas of BMN~\cite{Berenstein:2002jq}
a spin chain language for symmetric product orbifolds was put forward by several
authors~\cite{Gomis:2002qi,Lunin:2002fw,Gava:2002xb}.  
In~\cite{David:2008yk} (see also~\cite{Lee:2008sk}) the giant magnon solutions on the gravity side  were mapped
to certain   excitations above the chiral vacuum
in the symmetric product orbifold. A dynamical spin chain
picture based on the symmetry properties of the theory was suggested and
an all-order dispersion relation for the  magnon proposed
 using a central extension of the symmetry algebra.
However the magnon S-matrix has not been computed yet. While on the sigma model side this is in principle
a straightforward calculation, on the $CFT_2$ side even setting up the question is non-trivial
and seems to require  the proper construction of a spin chain in ``position space'', which would allow a concrete 
definition of asymptotic magnon excitations.

In this paper, which builds upon our recent  work \cite{Pakman:2009zz,Pakman:2009ab}, 
 we set  the stage for a systematic discussion of ``gauge invariant'' states in symmetric product orbifolds.
 We will study a specific ``position space''  spin chain picture for single-cycle operators,
 which are analogous to single-trace operators in gauge theory.
 We start in section \ref{defsec} by reviewing
basic facts about  ${\rm Sym}^N \, T^4$.
We  discuss the most generic gauge invariant operators and introduce a spin chain interpretation for single-cycle
operators, which can be regarded as  the elementary building blocks at large $N$.  
 In section \ref{treesec}  we define several different ways to introduce  ``impurities'' on the chain and 
 compute  two point functions of states with impurities  at the orbifold point. By analogy
 with the standard gauge theory case we will refer to these as 
``tree level'' calculations, since they are the closest we can get to  free field theory calculations. The analogy is however far from perfect,
since even at the orbifold point  correlators of twisted fields are non-trivial (there is no simple sense in which Wick theorem applies).
 Unlike the gauge theory spin chain, we encounter  large mixing
already at ``tree level'':  for example two states that differ by the position of one impurity
are in general not orthogonal, even at large $N$. 
In section \ref{1loopsec} we turn on an exactly marginal deformation that preserves $(4,4)$ supersymmetry
and discuss computations at leading non-trivial order in the
deformation  (``one loop''). Because of the complication of tree level mixing, 
 the fundamental question of whether the one-loop Hamiltonian is ``local'' is 
 difficult to answer. We find however some encouraging hints. 
In section \ref{sumsec} we summarize and discuss our results.
Several technical appendices complement the text.

We end this introduction with a brief recapitulation of the 
 current  evidence for the
  holographic correspondence between ${\rm Sym}^N \, T^4$
and type  IIB string theory on $AdS_3 \times S^3 \times T^4$~\cite{Maldacena:1997re}.
 See
\cite{Aharony:1999ti, Dijkgraaf:2000vr,David:2002wn, Martinec} for reviews.
The early checks of this duality included comparison of the moduli spaces \cite{Dijkgraaf:1998gf, Larsen:1999uk},
the spectra of both theories~\cite{Maldacena:1998bw, deBoer:1998ip, Kutasov:1998zh, Argurio:2000tb},
and the symmetries~\cite{Giveon:1998ns,Giveon:2003ku,Ashok:2009jw}.
Recently much progress was made in comparing
correlation functions. The structure constants of single-cycle operators in the chiral ring of the symmetric product
were computed early on in~\cite{Jevicki:1998bm} and, for a subset of these operators
 they were extended in~\cite{Lunin:2000yv, Lunin:2001pw} to the full 1/2 BPS $SU(2)$ multiplet. These three-point functions were exactly reproduced
in the string theory/supergravity
dual~\cite{Gaberdiel:2007vu, Dabholkar:2007ey, Pakman:2007hn, Taylor:2007hs} (see also~\cite{Giribet:2007wp,  Giribet:2008yt, Cardona:2009hk}),
which also predicts some correlators not yet computed in the symmetric product~\cite{Pakman:2007hn}.
The bulk-boundary agreement of these 
computations, performed far apart in the moduli space~\cite{Dijkgraaf:1998gf,Larsen:1999uk}, 
is explained by a non-renormalization theorem proved in~\cite{deBoer:2008ss}. The latter 
also holds for extremal correlators, a large  class of which was computed in the symmetric product orbifold in~\cite{Pakman:2009ab}.
For examples of explicit computations of correlators in symmetric product orbifolds see~\cite{Arutyunov:1997gt,Arutyunov:1997gi,Lunin:2000yv,Lunin:2001pw,Pakman:2009zz,Avery:2009tu,Avery:2009xr}.
The $AdS_3/CFT_2$ duality was also discussed in the pp-wave limit~\cite{Gomis:2002qi,Lunin:2002fw,Gava:2002xb,Hikida:2002in}.

\section{Definition of the spin chain}\label{defsec}

\subsection{Generic gauge invariant state}
We are interested in classifying and studying gauge invariant operators in symmetric product orbifolds.
For a general discussion of symmetric product orbifolds  we refer the reader to~\cite{Pakman:2009zz} and references therein.
The specific theory of our interest  will be ${\rm Sym}^N \, T^4$.
This theory has the following matter content: $4$ real left/right mover fermions and $4$ real bosons,
each coming in $N$ copies. The different copies of the fields are identified under the action of the  group of permutations $S_N$.
In  analogy to gauge theory we will refer to  the  index $I$ of the copies of the fields as the ``color'' index. 

The four real holomorphic fermions of $T^4$ can be combined,
in each copy $I$, into two complex fermions $\psi^1_I, \psi^2_I$, (with $I=1\dots N$) and bosonized as
\be
\psi_I^1&=& e^{i\phi_I^1}\,,
\\
\psi_I^2&=& e^{i\phi_I^2}\,, \qquad \qquad I=1,\ldots, N\,.
\ee
In each copy $I$, we pair the four real bosons into two complex bosons $X_\a^I$ ($a=1,2$).

The basic observables of a symmetric product orbifold  are the
twist fields~${\s}_{[g]}$, labeled by a conjugacy class $[g]$ of the permutation group.
``Gauge invariant'' twist fields~$\s_{[g]}$ can be constructed from ``gauge non-invariant'' ones, $\s_g$, associated
to a group element $g\in S_N$ and not to a conjugacy class. 
The operator $\s_g(z,\bar z)$ is defined as a ``defect'' imposing
 the following monodromies on the
different copies of the fields
\be
X^i_I(e^{2\pi i}\,z)\s_g(0)= X^i_{g(I)}( z)\s_g(0) \,,
\ee
and similarly   for the fermionic fields.
Gauge invariant  operators  are obtained by  averaging over the group orbit,
\be
\label{gaugeinvariant}
\s_{[g]} \equiv
{\mathcal A}_{[g]}(N)\,
 \sum_{h\in S(N)}\s_{h^{-1}\,g\,h} \, ,
\ee where ${\mathcal A}_{[g]}(N)$ is an appropriate normalization, see {\it e.g.}~\cite{Pakman:2009zz}.

The theory has  ${\cal N}=4$ supersymmetry. We will review the realization of the algebra
in terms of the fields of the theory in section \ref{susy}. A notable  set
of gauge invariant states is the set
of  (anti)chiral  states under some ${\mathcal N}=2$ subalgebra. Let us discuss these
first (see {\it e.g.}~\cite{Jevicki:1998bm,Lunin:2001pw,Pakman:2009ab}).
The $U(1)$ current of a ${\cal N}=2$ subalgebra of supersymmetry we will use is
\be\label{Rcurr}
J&=& \half\sum_{I=1}^N\left(\psi^1_I\,\psi^{1\,\dagger}_I+\psi^2_I\,\psi^{2\,\dagger}_I\right)
=\frac{i}{2}  \sum_{I=1}^N \left(\d \phi_I^1 + \d \phi_I^2\right)\,.
\ee
We define the gauge-non-invariant chiral operators associated to the single-cycle $g=(12\ldots n)$,
\be
o_{(12\ldots n)}^{(0,0)} &=&  e^{ i \frac{n-1}{2n} \sum_{I=1}^{n} (\phi_I^1 +  \phi_I^2  + \bar{\phi}_I^1 + \bar{\phi}_I^2 )}   {\sigma}_{(12\ldots n)}\, ,
\label{gnop1}
\\
o_{(12\ldots n)}^{(a=1,\bar{a}=1)} &=&  e^{ i \frac{n+1}{2n} \sum_{I=1}^{n} (\phi_I^1 + \bar{\phi}_I^1 ) + i \frac{n-1}{2n} \sum_{I=1}^{n}(\phi_I^2  +\bar{\phi}_I^2)}  {\sigma}_{(12\ldots n)}\, ,
\\
o_{(12\ldots n)}^{(a=2,\bar{a}=2)} &=&  e^{ i \frac{n-1}{2n} \sum_{I=1}^{n} (\phi_I^1 + \bar{\phi}_I^1 ) + i \frac{n+1}{2n} \sum_{I=1}^{n}(\phi_I^2  + \bar{\phi}_I^2)} {\sigma}_{(12\ldots n)}\, ,
\\
o_{(12\ldots n)}^{(2,2)} &=&  e^{ i \frac{n+1}{2n} \sum_{I=1}^{n} (\phi_I^1 +  \phi_I^2  +\bar{\phi}_I^1 + \bar{\phi}_I^2)}  {\sigma}_{(12\ldots n)}\,.
\label{gnop4}
\ee
The gauge invariant operators are obtained by summing over the group orbit,
\be
O_n^{(0,0)} &=&  \frac{1}{\sqrt{n\,N!(N-n)!}} \sum_{h\in S(N)}   o_{h^{-1} (12\ldots n)h}^{(0,0)}
\, ,
\label{op1}
\\
O_n^{(a,\bar{a})} &=&  \frac{1}{\sqrt{n\,N!(N-n)!}} \sum_{h\in S(N)}   o_{h^{-1} (12\ldots n)h}^{(a,\bar{a})}\, ,
\\
O_n^{(2,2)} &=&  \frac{1}{\sqrt{n\,N!(N-n)!}} \sum_{h\in S(N)}   o_{h^{-1} (12\ldots n)h}^{(2,2)} \,.
\label{op4}
\ee
The conformal dimensions  and charges are
\be
\Delta^{0}_n &=& Q^{0}_n = \frac{n -1 }{2}\, ,
\label{jepsn}
\\
\Delta^{a}_n &=& Q^{a}_n = \frac{n}{2}\, ,
\\
\Delta^{2}_n &=& Q^{2}_n = \frac{n +1 }{2}\, ,
\ee
and similarly for the antiholomorphic sector. The antichiral operators
$O_n^{(0,0) \dagger}, O_n^{(a,\bar{a}) \dagger}, O_n^{(2,2) \dagger}$
are obtained by reversing the sign in the exponents in (\ref{gnop1})-(\ref{gnop4}).\footnote{
One can also define operators with different left and right properties.} 

\

Let us now build a generic gauge invariant state.
We will denote schematically all the fields in copy $I$ as
$\chi_I$. 
The generic state invariant under the action of the permutation group $S_N$ is built from
\be
o_g={\mathcal G}\[\chi_I|\,g\cdot I = I\]\;{\mathcal F}\[\chi_I|\,g\cdot I \neq I\]\;\s_{g}.
\ee Here $g$ is some group element of $S_N$. We will refer to the twist field $\s_g$ as
 {\it bare} twist field to emphasize that it is an operator  without any ${\mathcal F}$ and ${\mathcal G}$  dressing. 
 The function ${\mathcal G}\[\chi_I|\,g\cdot I = I\]$
 commutes with the bare twist field by definition and is arbitrary. However, we have to demand that ${\mathcal F}\[\chi_I|\,g\cdot I \neq I\]$
 also commutes with $\s_g$ and this implies certain non trivial restrictions. In the example of chiral fields
 \eqref{gnop1}-\eqref{gnop4} the fermionic dressing is the ${\mathcal F}$ part and there is
 no ${\mathcal G}$ part. A set of gauge invariant states is then given by
\be\label{mostgen}
{\cal O}_{[g]}=\sum_{\beta\in S_N}\, o_{\beta g\beta^{-1}},
\ee 
and the most generic gauge invariant state is built from linear combinations of these states. The ultimate goal
is to find the spectrum of conformal dimensions of the gauge invariant operators after
 turning on interaction terms. While the general problem is hugely complicated, it should
 simplify in the large $N$ limit, which corresponds to the classical limit  on the string theory side.
The theory is expected to be integrable in this limit.

The states ${\cal O}_{[g]}$ are  classified according to the conjugacy class $[g]$, which is
specified by the cycle structure of the group element, {\it i.e.} the number of single-cycles and their length.
States with different cycle structure are exactly orthogonal, even at finite $N$,
since for $\langle o_g\, o_h\rangle$ to be 
non zero the product $g\,h$ has to be the identity.
It is common (see {\it e.g.}~\cite{Gomis:2002qi,Lunin:2002fw,David:2008yk})
to draw an analogy between  multi-cycle states in the symmetric product orbifold and
multi-trace states in gauge theories. Both, single traces and single-cycles, have a natural cyclic structure and are bounded in size
by the number of colors $N$. (In the symmetric product this is a strict bound, while in a gauge theory
it is the statement that a single trace operator longer than $N$ can be re-written as a linear combination
of shorter multi-traces.)
An important rationale for the analogy is  the fact that in both cases single (multiple) trace/cycle states correspond to
single (multiple) gravitons. This is confirmed  for single gravitons by the bulk-boundary agreement of three-point functions in both cases~\cite{Lee:1998bx,
Gaberdiel:2007vu, Dabholkar:2007ey, Pakman:2007hn, Taylor:2007hs}. While
multi-trace states in the gauge theory are not exactly orthogonal at finite $N$ (unlike multi-cycle states in the symmetric product),
they become orthogonal  at large $N$ (if the length of the traces is kept fixed), which is  the relevant limit for our discussion.

\begin{figure}[thbp]
\begin{center}
 \epsfig{file=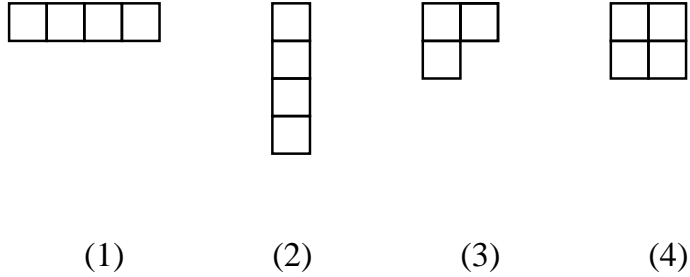,scale=0.5}
\end{center}
 \begin{center}
\caption{{\bf (1)} Young tableau representation of a state in the untwisted sector, ${\cal O}=\sum_{I\neq J\neq K\neq L}\,e^{i\left(\chi_I+\chi_J+\chi_K+\chi_L\right)}$.
Here ${\cal O}$ has only a ${\mathcal G}$ part.
{\bf (2)} Representation of a state in a twisted sector, ${\cal O}=\sum_{I\neq J\neq K\neq L}\,e^{i\left(\chi_I+\chi_J+\chi_K+\chi_L\right)}\s_{(I\, J\, K\, L)}$.
Here there is only an ${\mathcal F}$ part.
{\bf (3)} Representation of the state  ${\cal O}=\sum_{I\neq J\neq K}\,e^{i\left(\chi_I+\chi_J+\chi_K\right)}\s_{(I\, J)}$. The ${\mathcal G}$
dressing is ${\mathcal G}=e^{i\chi_K}$, and the ${\mathcal F}$ dressing is ${\mathcal F}=e^{i\left(\chi_I+\chi_J\right)}$.
{\bf (4)} Representation of the state  ${\cal O}=\sum_{I\neq J\neq K\neq L}\,e^{i\left(\chi_I+\chi_J+\chi_K+\chi_L\right)}\s_{(I\, J)}\s_{( K\, L)}$.
There is only ${\mathcal F}$ dressing, ${\mathcal F} = e^{i\left(\chi_I+\chi_J+\chi_K+\chi_L\right)}$.
 } \label{youngs}
\end{center}
\end{figure}

Since the cycle structure of $[g]$ is specified by a partition of $n$, 
with $n \leq N$,
we can conveniently associate $[g]$ to a Young tableau. If $n = \sum_k n_k$
where $n_k$ is the number of cycles of length $k$, we draw a tableau
with $n_k$ columns of length $k$, that is we associate
to each  single-cycle of length $k$  a column with $k$ boxes.
The untwisted sector states of the orbifold correspond to columns of the Young tableaux with a single box, and therefore
we can represent an untwisted state  which involves $k$ colors as a single row with $k$ boxes. 
This is illustrated in figure \ref{youngs}.
The total number $n$ of boxes in the Young tableaux is bounded by $N$, and this is related to the stringy exclusion principle~\cite{Maldacena:1998bw}.
We can use a similar Young tableaux representation for multi-trace states in a gauge theory, associating single-traces of length $k$ to columns of length $k$,
in keeping with our analogy between single-cycles and single-traces.
A column with a single box corresponds to the trace of a single field in the adjoint of $U(N)$, {\it i.e.} to its $U(1)$ part.
 In some rough sense the untwisted states of the symmetric product are analogous to states in the $U(1)$  decoupled sector of the gauge theory. 
 Unlike the symmetric product case, in the gauge theory
there is no upper bound on the total number of boxes.\footnote{
At finite $N$,  a useful orthogonal basis
for Yang-Mills operators built out of a single adjoint scalar $Z$
 is given by  the Schur polynomial basis~\cite{Corley:2001zk}. Schur polynomials are also naturally represented by Young tableaux,
which should  not be confused with the way we use Young tableaux to represent multi traces.
In the Schur basis,  a column of the Young tableau of length $k \lesssim N$ 
 is associated to a subdeterminant and is
holographic to  sphere giant  gravitons~\cite{McGreevy:2000cw}, while a row is holographic to
 AdS  giant gravitons~\cite{Hashimoto:2000zp}. See~\cite{Lunin:2002bj,Raju:2007uj} 
for  the issue of giant gravitons   in $AdS_3 \times S^3 \times T^4$.}

In what follows we consider the strict infinite $N$ limit of the symmetric product.  In this limit
correlators of generic gauge invariant operators factorize into correlators  of single-cycle
operators with trivial  ${\mathcal G}\[\chi_I\]$ part, and untwisted correlators. 
Single-cycle operators with trivial ${\mathcal G}\[\chi_I\]$ are
elementary building blocks, and at large $N$ form a close sector
 under the action of the dilatation operator, even away from the orbifold point.
They are expected to map to single closed string states on the dual side.

\subsection{Towards a spin chain}

As familiar, in a gauge theory we can represent single-trace operators
as spin chains with periodic boundary conditions. The simplest
example is the $SU(2)$ sector of ${\cal N} = 4$ SYM, which consists
of operators made of two adjoint scalars, $Z$ and $X$, which are thought as spin up and spin down.
 The  operator with only $Z$s, $\Tr Z^n$, is the vacuum of the spin chain (of length $n$); replacing $Z$ with $X$ at various sites we introduce ``impurities''.
 In some special gauge theories, {\it e.g.} ${\mathcal N}=4$ SYM~\cite{Minahan:2002ve},
the one-loop dilatation operator turns out to be the nearest-neighbor Hamiltonian of an integrable spin chain -- for
example, it is the Heisenberg spin chain in the $SU(2)$ sector. At higher orders
the Hamiltonian becomes very complicated, and it has proved more fruitful 
to phrase the integrability structure in terms of the S-matrix of asymptotic magnon excitations of the spin chain.

We would like to uncover the analogous notions for symmetric product orbifolds in general and in particular
for ${\rm Sym}^N \, T^4$.
 We assume the  analogy between a single-trace operator in a gauge theory and a single-cycle twist field in the symmetric product.
 Next we should specify what  we mean by an individual ``site'' of the single-cycle twist field.
 In Yang-Mills  the sites are identified with the elementary adjoint fields of the composite operator. In the symmetric product, it was suggested
by several authors~\cite{Gomis:2002qi,Lunin:2002fw,David:2008yk} to decompose a  given $n$-cycle
element of $S_N$   as a product of transpositions, and to consider this decomposition as a collection of $n-1$ sites.
For example
\be\label{transp}
(1\,2\,3\,4\,5) &=& (1\,5)(5\,4)(4\,3)(3\,2),
\\
&=&  (2\,3)(3\,4)(4\,5)(5\,1)\,.
\ee
There are however some  difficulties with this identification. First, as we sprinkle the different
sites  with  impurities,   the dressed transpositions will generally have singularities in the
OPE with one another, and some prescription must be specified
 in recombining the transpositions into a single-cycle.
More fundamentally, as is clear from the above example, the decomposition of a single-cycle into transpositions is not unique and it breaks the explicit cyclic structure of the single-cycle.
  Given a single-cycle operator  there is no canonical gauge invariant way to specify the sites of the  associated spin chain.

To avoid  these problems we propose to  identify the sites of the spin chain with
the  ``colors'' permuted by the cycle. This gives  the sites a natural cyclic ordering.
There {\it is} a natural gauge invariant way to act on a single site of the chain, as follows (see~\cite{David:2008yk} for similar manipulations).
We insert an impurity at a site $I$ by inserting an operator $\hat P_I$ that depends on the elementary
fields $\chi_I$ of that site (=color $I$),
\be
\hat P_I \,   \sigma_{(12 \dots n)} \, ,  
\ee
where $\sigma_n$ is the gauge non-invariant twist field. Then we 
 sum cyclically over  the $n$ sites of the chain,
 \be
\hat P \sigma_{(12\dots n)} \equiv \sum_{I=1}^n \hat P_I \,   \sigma_{(1 2 \dots n)} \, ,  
 \ee
 and finally we sum over all relabelings of the colors to get the desired gauge invariant operator
(as in~\eqref{mostgen}),
 \be
\sum_{h \in S_N}   \sum_{I=1}^n \hat P_I \,    \sigma_{h(1 2 \dots n)h^{-1}} \, .
 \ee
 In many concrete cases it will be useful to represent $\hat P_I$ as a contour integral
\bea
\hat P_I & = & \oint\frac{dz}{2\pi i} P_I(z) \\
\hat P\, \tilde o_n(0)&  \equiv& \sum_{I=1}^n\,\oint\frac{dz}{2\pi i} P_I(z)\, \tilde o_n(0)\,.
\eea 
Here $\tilde o_n(z)$ is a generic gauge non-invariant operator in the twisted sector associated to the single-cycle $(12\dots n)$.
It is most convenient to discuss symmetric product orbifolds on a covering surface where the twist fields disappear and the
fields are single-valued. The exact covering map, $z(t)$, is determined by the correlator being evaluated. The covering map has the property that near the location of single-cycle twist field, say $z=0$ and $t=0$, it behaves as $z(t)\sim a\, t^n$.
On the covering surface the action of an operator on a single color can be elegantly written as
\be	
\hat P\, \tilde o_n(0)\sim \oint\frac{dt}{2\pi i} \left(\frac{dt}{dz}\right)^{\delta-1}\,P(t)\, {\tilde o}_n^{(t)}(0).
\ee where $\delta$ is the conformal dimension of $P(z)$ and the superscript $t$ in $ {\tilde o}_n^{(t)}$ denotes that the operator is lifted to the covering surface, on which the bare twist field $\sigma_{n}$ disappears. The fact that we can easily lift the definition of the impurities
 to the covering surface will give us powerful computational tools to evaluate correlators.

To introduce two impurities at different sites of the chain one proceeds as follows.
The action of operator ${\hat P}^{(1)}$ and ${\hat P}^{(2)}$ on two different sites, $I$ and $I+L$, becomes
\be
\sum_{I=1}^n\,\oint\frac{dz}{2\pi i} P^{(1)}_I(z)P^{(2)}_{I+L}(z)\; \tilde o_n\to\oint\frac{dt}{2\pi i}
\left(\frac{dt}{dz}\right)^{\delta_1-1}\,
\left(\frac{dt_L}{dz}\right)^{\delta_2}\,
P^{(1)}(t)\,P^{(2)}(t_L(t))\, {\tilde o}_n^{(t)}.\nonumber\\
\ee  where the ordering on the color is defined by the cyclic ordering inferred from the structure of the twist field.
The function $t_L(t)$ satisfies the following
\be
z(t)=z(t_L(t)),
\ee and for $t$ in the vicinity of the insertion of the chiral field $t=0$ we have
\be
t_L(t)\sim e^{\frac{2\pi i }{n}\,L}\, t.
\ee Near the pre-image of a twist field on the covering surface if a position $t$ corresponds on the base to color $I$ then
the position $~e^{\frac{2\pi i }{n}\,L}\, t$ corresponds to color $I+L$. The function $t_L(t)$ is crucial to define the action
of operators on different sites of the chain and we will discuss several explicit examples in the following sections.

The generalization to states with many impurities is straightforward.
The most general single-cycle state with trivial ${\mathcal G}$ part is a linear combination of states of the form
\be
\hat o_n= \oint\frac{dt}{2\pi i}
\left(\frac{dt}{dz}\right)^{-1}
\prod_{I=1}^n\left(\frac{dt_{L=I}}{dz}\right)^{\delta_I}\prod_{I=1}^n\,P^{(I)}(t_{L=I})\;,
\ee with $t_{L=1}\equiv t$.

\begin{figure}[htbp]
\begin{center}
 \epsfig{file=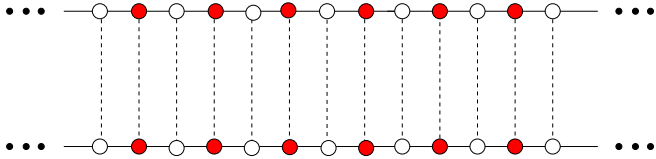,scale=0.2}
\end{center}
 \begin{center}
\caption{A qualitative picture of the two point diagram. The two solid lines correspond to  dressed twist field. We have taken the length
of the cycles to be large and have drawn the circles corresponding to twist fields as infinite lines.
  The diagrams of \cite{Pakman:2009zz} are obtained by identifying the vertices according to the dashed lines.
 } \label{2point}
\end{center}
\end{figure}

\
\begin{figure}[tbp]
\begin{center}
 \epsfig{file=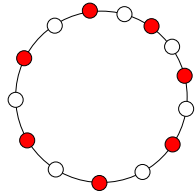,scale=0.2}
\end{center}
 \begin{center}
\caption{A vertex corresponding to a seven-cycle operator ${\cal O}_7$. The red circles are color loops.
 We can think of  the red dots as sites of a spin chain.   
 } \label{colorchain}
\end{center}
\end{figure}
We can represent the spin chain graphically, using a diagrammatic language for symmetric product orbifolds  introduced in \cite{Pakman:2009zz}
and briefly reviewed in appendix \ref{diagssec}.
In this language  a single-cycle twist field is  represented  as a loop with $2n$ vertices, see figure \ref{colorchain}.
The vertices are of two types, ``color'' and ``non-color''  vertices.
In the limit of large length of the chain we will depict it as a horizontal line. The ``Feynman diagrams'' for correlators are obtained by gluing together appropriately the
vertices of loops corresponding to different twist fields. An example of a tree level two point function is depicted in figure \ref{2point}.

\

\

The next step is to settle on a choice of vacuum for the spin chain
and to identify its  basic excitations, {\it i.e.} the impurities. A natural choice is to identify the vacuum as
the chiral state  with lowest dimension for a given length $n$~\cite{Gomis:2002qi,Lunin:2002fw,David:2008yk},
namely the state
\be
o_n\equiv o^{(0,0)}_n.
\ee
The other chiral primary
 fields \eqref{op1}-\eqref{op4} can be obtained from $o_n$ by the  action of some operators.
The chiral state satisfies $\Delta-J=0$ and the basic impurities can be defined as operators acting on a single site of a chain
and having minimal $\Delta-J$. In the next section we will discuss such excitations.
After the action of the impurities on $o_n$, we should symmetrize over the orbit of the $S_N$  group as in (\ref{op1}).

Before going into a detailed analysis of  the different types of impurities,
 let us  briefly review  the situation on the dual string side~\cite{Berenstein:2002jq,Gomis:2002qi,Lunin:2002fw,Gava:2002xb}.
In the light-cone gauge, the asymptotic excitations  of the (massive) worldsheet sigma model fall into two classes: there
are excitations related to  the $AdS_3\times S^3$ portion of the geometry, which are in one-to-one correspondence
with the generators of the superconformal algebra left unbroken by the light-cone gauge choice,
and excitations related to the $T^4$.
By contrast in  $AdS_5\times S^5$ 
 all  excitations are of the first type -- they are  associated with the unbroken superconformal generators.
We will indeed find that in ${\rm Sym}^N \, T^4$   the spectrum of  impurities  is in a sense richer than in  ${\mathcal N}=4$ SYM:
besides ``universal'' impurities associated with the symmetry generators, there are impurities associated to non-trivial primaries of $T^4$, 
which are perhaps more elementary.  In the CFT$_2$
the holomorphic objects with  lowest dimension are the fermions
($\Delta=\half$), while the symmetry generators  start at $\Delta =1$ (the $R$-symmetry currents).
Single-fermion impurities   in the spin chain of ${\rm Sym}^N \, T^4$ 
have no direct analogue in ${\cal N} = 4$ SYM.

\section{Impurities and tree level computations}\label{treesec}

In this section we consider different types of impurities  on the vacuum represented by the chiral operator $o_n$.
A natural choice is  to create impurities by acting with symmetry generators on the sites of the spin
 chain~\cite{Gomis:2002qi,Lunin:2002fw,David:2008yk}, {\it e.g.} the modes of the $R$-current and the susy generators.
However,  the smallest-dimensional fields in the orbifold are the fermions and thus the most fundamental impurities
one can build include these objects.\footnote{One can always restrict to states with impurities generated only by symmetry currents, as these
form a closed sector of the theory.}
In what follows we will study impurities generated by modes of the fermions ($\psi^a$ and $\psi^{a\,\dagger}$), and by modes of the bosons ($\d X^i$).
We also discuss impurities generated by modes of the $R$-current $J^-$, which is a quadratic composite of the fermions.  
The discussion will focus mostly on states with two impurities at a distance~$L$, and in each case we check whether states with different $L$'s
are orthogonal at tree level  - as is the case in ${\mathcal N}=4$ SYM in the limit of large $N$.
 In symmetric product CFTs at tree level 
the only covering surfaces contributing to two point functions have topology of a sphere.\footnote{This can be deduced  for instance from 
the Riemann-Hurwitz relation determining the genus of a covering surface by noting that the number of colors is uniquely determined
to be equal to the size of the cycles, see {\it e.g.}~\cite{Pakman:2009zz}.}
 Thus,  the full $N$ dependence is given by a simple over-all combinatorial factor~\cite{Pakman:2009zz}. In particular the
 large $N$ considerations do not play any role 
in tree level computations. 
As we will see, some types of two-impurity states are orthogonal, some are not and some are orthogonal only when the length $n$ of the spin chain (not to be confused
with the number  $N$) is taken to be large. 
These results illustrate the lack of a sharp analogy between the symmetric product and ${\mathcal N}=4$ SYM.

\subsection{Fermionic Impurities}
We first introduce fermionic impurities by directly constructing  the states  in the bosonized language for the fermions.
We then show that this construction can be recast as the action of a certain current algebra on the chiral vacuum.

\subsection*{Direct Construction}

The states with a single impurity are defined as the following currents acting on the chiral vacuum~$o_n$,
\be
A^a &=& \sum_{I=1}^n\oint\frac{dz}{2\pi i} \frac{1}{z}\,e^{-i\phi^a_I(z)} \,, \qquad a=1,2
\\
&=& \sum_{I=1}^n \psi^a_{I,-1/2} \,,
\\
B^a &=&\sum_{I=1}^n\oint\frac{dz}{2\pi i} \frac{1}{z^2}\,e^{i\phi^{a}_I(z)} \,, \qquad a=1,2
\\
&=&  \sum_{I=1}^n \psi^{a \dagger}_{I,-3/2} \,,
\ee
and analogously for the anti holomorphic sector.  Note that these impurities have different conformal dimensions and thus are orthogonal at tree level. Strictly speaking this definition is correct only for odd $n$, as otherwise the fermions are antiperiodic when rotated around the twist field~\cite{Lunin:2002fw}.
The currents $A^a$ increase the dimension by $1/2$ and decrease the charge by $1/2$, while
the currents $B^a$ increase the dimension by $3/2$ and increase the charge by $1/2$, so both types of impurities
satisfy $\Delta-J=1$.

Note that we could also act on $o_n$ with the following current
\be
\sum_{I=1}^n\oint\frac{dz}{2\pi i} \frac{1}{z}\,e^{i\phi^{a}_I(z)}
= \sum_{I=1}^n \psi^{a \dagger}_{I,-1/2} \,,
\ee
which has $\Delta-J=0$. However, this gives us just the fermionic chiral state $o_n^{(a,0)}$, with 
$(\Delta,\,\bar \Delta)=(\frac{n}{2},\,\frac{n-1}{2})$, and thus we should not treat this as an impurity.

Let us consider the $A$ type impurities in more detail. The  current generating  two impurities
is defined by
\be
&& \sum_{I=1}^n\oint\frac{dz}{2\pi i} e^{-i\phi^a_I(z)}\,e^{-i\phi^b_{I+L}(z)}
\\
= && \sum_{I=1}^n \,\, [\psi_I^a \psi_{I+L}^b]_{0} \,,
\ee
where $[\ldots]_0$ denotes the zero mode in the mode expansion of the normal product inside the square brackets.
On the covering surface this becomes
\be
\oint\frac{dt}{2\pi i}\sqrt{\frac{\d t_L}{\d t}}\,(t-t_L)^{\delta_{a,b}}\, e^{-i\phi^a(t)-i\phi^b(t_L)}.
\ee
Note that this expression is single valued and well defined, as near $t=0$ the function $t_L(t)$ by definition  behaves as
 $t_L(t)\sim e^{\frac{2\pi i}{n}L}\, t$ and thus  we do not cross the branch cut of the square root. This impurity does not change the conformal dimension of
 the chiral operator but reduces its charge by one unit, and thus has $\Delta-J=1$.
 For $L>0$, this current creates a two-impurity state with minimal $\Delta-J$ and this is in
 contrast to ${\mathcal N}=4$ SYM where such states do not exist.

We define a state with two impurities by acting with the above current on a
a state $o_n$ to obtain\footnote{Omitting an overall rescaling factor which should be correctly regularized in any correlator, {\it e.g.} following Lunin-Mathur procedure~\cite{Lunin:2000yv}.}
\be
|L\rangle &=&  \oint\frac{dt}{2\pi i}\sqrt{\frac{\d t_L}{\d t}}\,(t-t_L)^{\delta_{a,b}}\, e^{-i\phi^a(t)-i\phi^b(t_L)} e^{i\frac{n-1}{2}(\phi^1(t)
+\phi^2(t))}e^{i\frac{n-1}{2}(\bar\phi^1(\bar t)
+\bar\phi^2(\bar t))} |0\rangle,
\label{Lstate}
\\
&=&\oint\frac{dt}{2\pi i}\sqrt{\frac{\d t_L}{\d t}}\,\frac{(t-t_L)^{\delta_{a,b}}}{t^{\frac{n-1}{2}}{t_L}^{\frac{n-1}{2}}}\,
 e^{-i\phi^a(t)-i\phi^b(t_L)+i\frac{n-1}{2}(\phi^1(t)
+\phi^2(t))}e^{i\frac{n-1}{2}(\bar\phi^1(\bar t)
+\bar\phi^2(\bar t))} |0\rangle\nonumber \,.
\ee
Note that taking $L=0$ for $a=b$ we get zero, but if $a\neq b$ we get single impurity created by~$J_0^-$.

The generalization to an operator creating an even number, $2l+2$, of  impurities, {\it i.e.} a bosonic chain, is
\be
\sum_{I=1}^n\oint\frac{dz}{2\pi i}\, z^{l} \prod_{j=1}^{2l+2} e^{-i\phi^{b_j}_{I+L_j}(z)},
\ee with $L_j<L_{j+1}$, and $L_1=0$.
On the covering surface this becomes
\be
\oint\frac{dt}{2\pi i}\, z(t)^l\,\sqrt{z'(t)}\prod_{j=2}^{2l+2}\frac{1}{\sqrt{z'(t_{L_j})}}\,\prod_{j\neq j'}(t_{L_j}-t_{L_{j'}})^{\delta_{b_j,b_j'}}\, e^{-i\sum_{j=1}^{2l+2} \phi^{b_j}(t_{L_j})}.
\ee These states have $\Delta-J=l+1$.

The generalization to an operator creating an odd number, $2l+1$, of impurities, {\it i.e.} a fermionic chain,  is
\be
\sum_{I=1}^n\oint\frac{dz}{2\pi i}\, z^{l-1} \prod_{j=1}^{2l+1} e^{-i\phi^{b_j}_{I+L_j}(z)},
\ee with $L_j<L_{j+1}$, and $L_1=0$.
On the covering surface this becomes
\be
\oint\frac{dt}{2\pi i}\, z(t)^{l-1}\,\sqrt{z'(t)}\prod_{j=2}^{2l+1}\frac{1}{\sqrt{z'(t_{L_j})}}\,\prod_{j\neq j'}(t_{L_j}-t_{L_{j'}})^{\delta_{b_j,b_j'}}\, e^{-i\sum_{j=1}^{2l+1} \phi^{b_j}(t_{L_j})}.
\ee This definition holds for odd $n$. These states have $\Delta-J=l+1$.
 We note that there are bosonic and fermionic chains with same $\Delta-J$. Note also that  $\Delta-J$  for the
chain is roughly twice the number of the impurities.

Let us compute two point functions of two-impurity states at tree level.
Putting the twist fields at $0$ and $\infty$, both in the base and the covering sphere, the map to the covering surface
is
\be \label{treemap}
z(t)=t^n.
\ee
In particular this implies that
\be
t_L(t)=e^{\frac{2\pi i}{n}\,L} \, t.
\ee
We define
\be
p=e^{\frac{2\pi i}{n}L},\qquad q=e^{\frac{2\pi i}{n}M},
\ee to obtain
\be
\langle M|L\rangle  \sim p\,(1-p)^{\delta_{ab}}(1-q)^{\delta_{a'b'}}\oint\frac{dt'}{2\pi i} \oint\frac{dt}{2\pi i}
\frac{t^{1-n+\delta_{ab}}\,{t'}^{n-1+\delta_{a'b'}-\delta_{ab'}-\delta_{a'b'}-\delta_{aa'}-\delta_{bb'}}}
{(1-\frac{t}{t'})^{\delta_{aa'}}(1-p\frac{t}{t'})^{\delta_{ba'}}
(q-\frac{t}{t'})^{\delta_{ab'}}(q-p\frac{t}{t'})^{\delta_{bb'}}}.\nonumber\\
\ee In order to get simple poles for both contour integrals above we have to satisfy
\be
2-n+\delta_{ab}+\a=0,\qquad n+\delta_{a'b'}-\delta_{ab'}-\delta_{a'b}-\delta_{aa'}-\delta_{bb'}-\a=0,
\ee
where $\a$ is the expansion order in $t/t'$ of the denominator.
This implies the selection rule
\be
\delta_{ab}+\delta_{a'b'}-\delta_{ab'}-\delta_{a'b}-\delta_{aa'}-\delta_{bb'}+2=0,
\ee
which  implies either  $a=b=a'=b'$ or $a=a'\neq b=b'$ ($a=b'\neq b=a'$). Evaluating the integrals by residues
we obtain, up to an overall constant, for the first case,
\be
\label{Lnorm}
\langle M|L\rangle\sim -n\,\delta_{L,M}+ n\,\delta_{L,-M},
\ee
and for the second case
\be
\langle M|L\rangle\sim -1+n\,\delta_{L,M},\qquad (\;\langle M|L\rangle\sim -1+n\,\delta_{L,-M}\;) \,.
\ee In the large $n$ limit, {\it i.e.} long spin chain, all these two-impurities states are orthogonal.

\

As another example of  tree level properties let us consider the two point functions of states with two $\psi_1$ and
two $\psi_2$ impurities. The two point function is given by
\be
&&\langle \{M_i\}|\{L_i\}\rangle  \sim\\
&&\oint\frac{dt'}{2\pi i} \oint\frac{dt}{2\pi i}
\frac{p\hat p_1\hat p_2\,(1-p)\,(\hat p_1-\hat p_2)(1-q)(\hat q_1-\hat q_2)\;t^{5-2n}\,{t'}^{2n-7}}
{(1-\frac{t}{t'})(1-p\frac{t}{t'})
(q-\frac{t}{t'})(q-p\frac{t}{t'})
(\hat q_1-\hat p_1\frac{t}{t'})(\hat q_1-\hat p_2\frac{t}{t'})
(\hat q_2-\hat p_1\frac{t}{t'})(\hat q_2-\hat p_2\frac{t}{t'})}.\nonumber
\ee
Here the $\psi_1$ impurities are at sites $e^{\frac{2\pi i}{n}L}=1,\, p$ and $e^{\frac{2\pi i}{n}M}=1,\,  q$. The
$\psi_2$ impurities are at  $e^{\frac{2\pi i}{n}L}= \hat p_1,\, \hat p_2$ and
 $e^{\frac{2\pi i}{n}M}= \hat q_1,\, \hat q_2$. For generic values of $M_i$ and $L_i$ the above two point function is
exactly zero. It is non vanishing when at least one of the $L_i$ and $M_i$ overlap. For instance if $p=q$ the above is equal to
\be
\frac{2 n \hat p_1 \hat p_2(\hat p_1 - \hat p_2)  (\hat q_1 - \hat q_2)}{(\hat p_1 - \hat q_1) (\hat q_1-\hat p_2 ) (\hat p_1 - \hat q_2)
 (\hat p_2 - \hat q_2)}.
\ee This expression scales as $n$. However, if say $\hat p_1$ is close enough to $\hat q_1$, {\it i.e.} $L_i-M_j\ll n$,
 then the scaling is enhanced to $n^2$. Farther, if all the impurities are ``approximately''
 aligned in the above sense then the behavior is enhanced to $n^3$.

  In general, we can say that the states with impurities are orthogonal until at least one pair of impurities aligns
on the two chains. Then, even in the large $n$ limit, if the rest of the impurities are approximately aligned, {\it i.e.} the difference in their position is much less than $n$, the two point function is not zero at tree level and the states mix. This fact has to be contrasted
with $YM$. There two states with impurities which are not aligned do not mix because of the large $N$ suppression. However, the large
$N$ limit in symmetric product orbifold does not rule out such mixings.

\subsection*{Current Algebra Construction}
In this subsection we present an algebraic approach to the impurities introduced above.
For simplicity, we focus on two-impurity states with $a=b$ and we omit the labels $a,b$ from the fermions.

The exponent that dresses the spin field of the chiral vacuum $o_n$ in (\ref{gnop1}) is invariant under $Z_n$ action  of
the twist field. In the same spirit, we define the following
 $Z_n$-invariant currents,
\be
\label{curr1}
P^{-}_{\! L}  &\equiv& \sum_{I=1}^n \psi^{\dagger}_I \psi^{\dagger}_{I+L} = \sum_{r \in \mathbb{Z}} \frac{P^{-}_{\! L,r}}{z^{m+1}},
\\
P^{+}_{\! L} &\equiv& \sum_{I=1}^n \psi_I \psi_{I+L} = \sum_{r \in \mathbb{Z}} \frac{P^{+}_{\! L,r}}{z^{m+1}},
\\
\label{curr2}
N_{L} &\equiv& \sum_{I=1}^n \psi_I \psi^{\dagger}_{I+L} + \psi_{I+L} \psi^{\dagger}_{I} = \sum_{r \in \mathbb{Z}} \frac{N_{\! L,r}}{z^{m+1}},
\ee
which have dimension $\Delta=1$ and charges $\pm 1$ and $0$ under the $U(1)$ current $J$.
Since these currents are $Z_n$-invariant they have integer-moded expansions near the chiral vacuum $o_n$,
even though the modes cannot be expressed easily in terms of the modes of the fermions $\psi_I$.
Note the properties
\be
P^{\pm}_{\! 0} &=& 0\,,
\\
P^{\pm}_{\! - L} &=& -P^{\pm}_{\! L}  \,,
\\
N_{L}&=&  N_{-L} =  N_L^{\dagger}   \,.
\label{ndag}
\ee
Using the same contour integral arguments as in the previous section, it is easy to verify that the chiral vacuum
\be
|n \rangle  \equiv o_n  |0\rangle
\ee
is a highest weight state for the currents $P^{\pm}_{\! L}, N_{L}$,
satisfying
\be
P^{+}_{\! L, m}|n \rangle  &=& 0 \qquad \qquad m \geq - 1\,,
\\
P^{-}_{\! L, m}|n \rangle &=& 0 \qquad \qquad m \geq 1\,,
\\
N_{L, m}|n \rangle &=& 0 \qquad \qquad m \geq 1\,,
\\
N_{L,0} |n \rangle &=& c_L |n \rangle\,,
\ee
with
\be
c_L &=& 1 -n\delta_{L,0}\,.
\ee
In this language, the two-impurities states (\ref{Lstate}) are given by
\be
|L\rangle = P_{\! L,0}^{-}|n\rangle \,,
\label{2imp}
\ee
and acting with higher modes we can build all the module of the algebra.
Using
\be
\psi_i(z) \psi_j^{\dagger}(w) \sim \frac{\delta_{ij}}{z-w}\,,
\ee
we can obtain the algebra satisfied by the currents,
\be
P^{+}_{\! L_1} P^{-}_{\! L_2} \sim
 \frac{ \delta_{L_1,L_2} n}{(z-w)^2}  +  \frac{N_{L_1+L_2} - N_{L_1-L_2}}{z-w}\,, \qquad \qquad \qquad
\ee
or equivalently
\be
\[ P^{+}_{\! L_1, r} P^{-}_{\!  L_2, s} \] &=& n r \delta_{L_1,L_2} \delta_{r+s,0}
+  N_{L_2+L_1, r+s} - N_{-L_2-L_1, r+s} \,.
\label{alg}
\ee
Using now these commutation relations, we can compute the norm
\be
\langle L_1 |L_2  \rangle &=&
\langle n |P^{+}_{L_1,0}P^{-}_{L_2,0}  |n  \rangle
\\
&=&   \langle n |N_{L_2+L_1,0}  - N_{-L_2+L_1,0}  |n  \rangle
\\
&=& c_{L_2+L_1} - c_{L_2-L_1}
\\
&=& n \delta_{L_1,L_2} - n \delta_{L_1,-L_2}
\ee
which coincides, up to an overall constant which is not determined by the covering surface method, with (\ref{Lnorm}).
Impurities with $a\neq b$  can be obtained in a similar way, and the currents (\ref{curr1})-(\ref{curr2}) can be easily generalized  to the cases of three and more impurities. It would be interesting to compute the commutators for these cases and study the representation theory of these algebras.

\subsection{$J^-$ impurities}
In this section we discuss impurities built from modes of $J^-$.
Since the latter is a quadratic combination of the fermions, $\psi^1\psi^2$, the state with a single $J^-$ impurity is a special case of
the fermionic two-impurities state we discussed in the previous  section.\footnote{
More generally, the two fermions $\psi^a$ have same quantum numbers as the two supersymmetry generators
broken by the vacuum, the quadratic combination $\psi^1\psi^2$ is  the $R$-current $J^-$, and
the combination $\psi^1\psi^{\dagger\, 2}$ has the same quantum numbers as the derivative
 $\d$.}

But considering $J^-$ as the basic impurity, we can build a state with two or more impurities of type $J^-$ as follows.
Denoting by  $J^-_{(k)}$ the $J^-$ current acting on $k$th copy,
a two-impurities state is obtained  from the
zero mode of the combined operator (assuming that the current acts on state at $z=0$ for simplicity)
\be
\sum_{k=1}^n J_{(k)}^-J_{(k+L)}^-\to \sum_{k=1}^n\oint \frac{dz}{2\pi i}\, z\,
e^{-i\phi^1_k(z)-i\phi^2_k(z)}e^{-i\phi^1_{k+L}(z)-i\phi^2_{k+L}(z)}.
\ee
Lifting this to the covering surface we obtain (assuming that the image of the twist field on the covering surface is $t=0$)
\be
\sum_{k=1}^n J_{(k)}^-J_{(k+L)}^-&\to& \oint \frac{dt}{2\pi i}\, \frac{\d t_L}{\d z}\,z(t)
e^{-i\phi^1(t)-i\phi^2(t)}e^{-i\phi^1(t_L)-i\phi^2(t_L)}\\
&&=\oint \frac{dt}{2\pi i}\, \frac{\d t_L}{\d z}\,z(t)\,\left(t-t_L\right)^2
e^{-i\phi^1(t)-i\phi^2(t)-i\phi^1(t_L)-i\phi^2(t_L)}.\nonumber
\ee
The state is obtained by acting with this operator on the chiral vacuum  $o_n$ to obtain
\be
|L\rangle_{J^-}&=&\oint \frac{dt}{2\pi i}\,
 \frac{\frac{\d t_L}{\d z}\,z(t)\,\left(t-t_L\right)^2}{t^{n-1}\,t_L^{n-1}}\,
e^{-i\phi^1(t)-i\phi^2(t)-i\phi^1(t_L)-i\phi^2(t_L)+i\frac{n-1}{2}(\phi^1+\phi^2)}
e^{i\frac{n-1}{2}(\bar\phi^1+\bar\phi^2)}|0\rangle .
\nonumber
\\
\ee

Let us now compute the two point a two point function of the two-impurities states above.
Using the map to the covering surface \eqref{treemap} the two point function can be evaluated to be proportional to
\be
{}_{J^-}\langle M|L\rangle_{J^-}\sim\frac{1}{n^2}\oint \frac{dt}{2\pi i}\oint \frac{dt'}{2\pi i}\frac{p^2\,(1-p)^2\,(1-q)^2\,t^{5-2n}\,{(t')}^{2n-7}}
{(1-\frac{t}{t'})^2(1-p\frac{t}{t'})^2(q-\frac{t}{t'})^2(q-p\frac{t}{t'})^2},
\ee  where
$p=  e^{\frac{2\pi i }{n}\,L}$, $q=e^{\frac{2\pi i }{n}\,M}$ as before.
 Performing explicitly the computation one obtains
\be\label{diagmatrix1}
{}_{J^-}\langle M|L\rangle_{J^-}\sim
\left\{
\begin{array}{cc}
 L\neq \pm M & \frac{4}{n}-\frac{1}{n}\frac{1}{\sin^2\frac{\pi (L-M)}{n}}-\frac{1}{n}\frac{1}{\sin^2\frac{\pi (L+M)}{n}}\\
&\\
 L=\pm M & \frac{17 - 12 n + 4 n^2}{3n}-\frac{5}{4n}\frac{1}{\sin^2\frac{\pi L}{n}}-\frac{1}{4n}\frac{1}{\cos^2\frac{\pi L}{n}}\\
&\\
 L= \pm M=\pm\frac{n}{2} & \frac{16(2 - 3 n +  n^2)}{3n}\\
\end{array}
\right.
\ee
Note thus that our states are not orthogonal at tree level. In appendix \ref{diagapp}
we comment on the diagonalization of these states.

Let us consider the large $n$ limit. In this case if $L\neq\pm M$
the two point function scales as $n^{-1}$. However,  the $L=\pm M$ case scales as $n$. Thus, naively
in the large $n$ limit the states with impurities are approximately orthogonal and the mixing matrix is
proportional to the identity with coefficient scaling as $n$. However note that if $L-M\ll n$ for instance the term
$\frac{1}{n}\frac{1}{\sin^2\frac{\pi (L-M)}{n}}$ will scale as $n$ and thus will not be subleading.
 Thus, even in the large $n$ limit chains with close $L$ and $M$ mix.

\subsection{Fractional moded $J^-$ impurities}

In the previous sections we defined a spin chain in  ``position space'' using the functions~$t_L(t)$, but we can also
define  states in  ``momentum space''. One  way to do so would be  to Fourier transform the states we defined above.
However, in orbifold theories there is a  natural definition using
using fractional modes of the fields. For concreteness, let us discuss here the fractional
modes of the $R$ current $J^-$.  In  the presence of twist $n$ field we define~\cite{Lunin:2001pw}
\be
J^-_{-m/n}(z)\equiv \oint\frac{dz}{2\pi i} \sum_{k=1}^n J^-_{(k)}\, e^{-\frac{2\pi i m(k-1)}{n}}\,z^{-\frac{m}{n}}.
\ee This operator is gauge invariant and has dimension $\frac{m}{n}$. Note that here $\Delta-J\neq 1$.
We can refer to the quantum number $m$ as a ``momentum'' variable.

 The fractional moded operators are lifted in a very simple manner to
the covering surface
\be
J^-_{-m/n}\to \oint \frac{dt}{2\pi i}\, z(t)^{-\frac{m}{n}}\, J^-(t).
\ee
Note that these operators on the covering surface become just the integer modes of $J^-$ as $z(t)\sim t^n$ near $t=0$.
We can act with these operators on the chiral primary $o_n$ to obtain a general state of the form
\be
\prod_{k=1}^s\, J^-_{-m_k/n}\, o_n.
\ee
Note that if $\sum_{k=1}^sm_k=0$ we have a state with charge shifted by $s$ and unshifted dimension, implying $\Delta-J=s$.

Let us consider a state with two impurities
\be
\hat o_k\equiv J^-_{k/n}\,J^-_{-k/n} o_n&=&\\ \oint \frac{dt'}{2\pi i}\oint \frac{dt}{2\pi i}\, \left(\frac{z(t)}{z(t')}\right)^{-\frac{k}{n}}&&
\frac{(t-t')^2}{t^{n-1}{t'}^{n-1}}\,e^{-i\phi^1(t)-i\phi^2(t)-i\phi^1(t')-i\phi^2(t')+i\frac{n-1}{2}(\phi^1+\phi^2)}
e^{i\frac{n-1}{2}(\bar\phi^1+\bar\phi^2)}.\nonumber
\ee Note that the ``momentum'' $k$ is bounded. The map near zero behaves as $z(t)\sim t^n$ and for $k\geq n-1$
the $t'$ integration does not have a pole and thus vanishes. For $k\leq 1-n$ the $t$ integration vanishes for the same reason. Of course we can take
$k$ to be non negative as $J^-_{k/n}$ commutes with $J^-_{-k/n}$ and thus the independent values of $k$ are $0,\dots ,n-2$.

Let us compute  two point function of states with two impurities,
\be
\langle\hat o_k | \hat o_m\rangle\sim&&\,\oint \frac{dt'}{2\pi i}\oint \frac{dt}{2\pi i}\,\oint \frac{ds'}{2\pi i}\oint \frac{ds}{2\pi i}\, \left(\frac{t}{t'}\right)^{-k}\left(\frac{s}{s'}\right)^{-m}
\frac{(t-t')^2}{t^{n-1}{t'}^{n-1}}\frac{(s-s')^2}{s^{1-n}{s'}^{1-n}}\nonumber\\
&&\frac{1}{(t-s)^2(t'-s')^2(t-s')^2(t'-s)^2}.
\ee We assume that the contours are ordered as $|t'|<|t|<|s'|<|s|$ and write
\be
\langle\hat o_k| \hat o_m\rangle\sim&&\,\oint \frac{dt'}{2\pi i}\oint \frac{dt}{2\pi i}\,\oint \frac{ds'}{2\pi i}\oint \frac{ds}{2\pi i}\,
\frac{{t'}^{k+1-n}t^{-k-n+3}(1-\frac{t'}{t})^2
s^{n-m-3}{s'}^{n+m-5}(1-\frac{s'}{s})^2}{(1-\frac{t}{s})^2(1-\frac{t'}{s'})^2(1-\frac{t}{s'})^2(1-\frac{t'}{s})^2}
\nonumber\\
 \ee This can be evaluated to give
\be
 \langle\hat o_k | \hat o_m\rangle\sim\left\{
\begin{array}{cc}
 k=m&\qquad (n-2)^2-(k-1)^2+\delta_{k,0}\,(n-1)^2\\
 k> m &\qquad -2\,(n-k-1)\\
 m> k &\qquad -2\,(n-m-1)\\
\end{array}
\right.
\ee We see that these operators also mix at tree level.  However, in the strict infinite $n$ limit
rescaling the operators with $1/n$ we get that this basis is orthogonal. To get an orthonormal basis we have to rescale
with $\frac{1}{n}\frac{1}{\sqrt{1-\frac{k^2}{n^2}}}$. Thus defining $p=k/n$  we get an orthonormal set of states
labeled by $p$ such that  $|p|-1$ is finite.

\

The operators in ``position space'' we defined in previous section using $t_L(t)$ can be obtained from states built using
fractional modes as follows.
Let us consider the following operator
\be
&&\hat o^{pos}_L\equiv\sum_{m=0}^\infty\,e^{\frac{2\pi i\, m\, L}{n}} \hat o_m
\to\\
&&\oint \frac{dt'}{2\pi i}\oint \frac{dt}{2\pi i}\, \frac{1}{1-e^{\frac{2\pi i\, L}{n}}\,\left(\frac{z(t')}{z(t)}\right)^{\frac{1}{n}}}
\frac{(t-t')^2}{t^{n-1}{t'}^{n-1}}\,e^{-i\phi^1(t)-i\phi^2(t)-i\phi^1(t')-i\phi^2(t')+i\frac{n-1}{2}(\phi^1+\phi^2)}
e^{i\frac{n-1}{2}(\bar\phi^1+\bar\phi^2)},\nonumber
\ee where we assumed that $|z(t)|>|z(t')|$. We have interchanged the order of taking the contour integral and performing the infinite sum.
These two limits do not commute.
 Note that if we were to truncate the sum above at finite $m$ the pole in the integrand coming from summing up the
 geometric series  would not have appeared.
This is a sign of the fact that $J^-_{k/n}$ commutes with $J^-_{-k/n}$. However, upon changing the limits
 we develop a pole and the two currents effectively cease to commute.
 The commutator term which we develop is exactly the operator we defined in the previous section.
We can deform the $t$ integration to two contours, one around $t'$ and the other around $0$.
 Thus, computing it with residue theorem we get
\be
 &&o^{pos}_L\equiv\hat o^{pos}_L+\sum_{m=-\infty}^{-1}\,e^{\frac{2\pi i\, m\, L}{n}} \hat o_m=	
\sum_{m=-\infty}^{\infty}\,e^{\frac{2\pi i\, m\, L}{n}} \hat o_m\to\\
&&\oint \frac{dt'}{2\pi i}\, z(t_L)\,\frac{d\, t_L}{d\,z}
\frac{(t_L-t')^2}{t_L^{n-1}{t'}^{n-1}}\,e^{-i\phi^1(t_L)-i\phi^2(t_L)-i\phi^1(t')-i\phi^2(t')+i\frac{n-1}{2}(\phi^1+\phi^2)}
e^{i\frac{n-1}{2}(\bar\phi^1+\bar\phi^2)},\nonumber
\ee where $t_L$ is defined by
\be
z(t_L)^{1/n}= e^{\frac{2\pi i\, L}{n}}\,z(t)^{1/n}.
\ee This is the operator we defined in the previous section. 

\subsection{$\d X$ impurities}
We can add impurities using the modes of $\d X^i$, $\d X^{\dagger i}$ and the antiholomorphic counterparts.
A chain with $l$ impurities is defined as
\be
\sum_{I=1}^n \oint\frac{dz}{2\pi i}\frac{1}{z} \prod_{j=1}^l \d X_{I+L_j}^{i_j} o_n.
\ee The bosons do not carry $R$-charge. However, they carry a charge under the outer automorphism $SU(2)$.
 As we will see in what follows the interactions are invariant under this $SU(2)$ and thus we can use it
as a selection rule. The above state has $\Delta-J=l$. On the covering surface it becomes
\be
\oint \frac{dt}{2\pi i}\frac{1}{z(t)}\prod_{k=2}^l\frac{\d t_{L_j}}{\d z}\, \prod_{j=1}^l \d X^{i_j}(t_{L_j}) o_n.
\ee
Let us consider the two point function in the simplest case  when all the bosons are of the same kind.
 This case is simple because  there are no contractions
between the $\d X$s of the chain.
 The tree level two point function is
\be
&&\oint\frac{dt}{2\pi i}\oint\frac{dt'}{2\pi i}z(t')^{2l-1}\prod_{k=2}^l\frac{\d t'_{M_j}}{\d z} \frac{1}{z(t)}\prod_{k=2}^l\frac{\d t_{L_j}}{\d z}
\sum_{\a\in S_l}\prod_{i=1}^l\frac{1}{(t_{L_i}-t'_{M_{\a(i)}})^2}=\\
&&\frac{1}{n^{2l-2}}\prod_{k=2}^l q_{l}p_{l}\oint\frac{dt}{2\pi i}\oint\frac{dt'}{2\pi i}{t'}^{n(2l-1)-(n-1)(l-1)-2l} t^{-n-(n-1)(l-1)}
\sum_{\a\in S_l}\prod_{i=1}^l\frac{1}{(p_i \frac{t}{t'}-q_{\a(i)})^2}\nonumber\\
&&\frac{1}{n^{2l-2}}\prod_{k=2}^l q_{l}p_{l}\oint\frac{dt}{2\pi i}\oint\frac{dt'}{2\pi i}{t'}^{nl-l-1} t^{-nl+l-1}
\sum_{\a\in S_l}\prod_{i=1}^l\frac{1}{(p_i \frac{t}{t'}-q_{\a(i)})^2}\,,\nonumber
\ee
where $S_l$ is the group of permutations of $l$ objects.
Taking a state with two impurities, {\it i.e.} $l=2$, we get that the two point function is
\be
{\mathcal I}_{L\, M}=
\left\{
\begin{array}{cc}
 L\neq \pm M & -\frac{1}{n}\frac{1}{\sin^2\frac{\pi (L-M)}{n}}-\frac{1}{n}\frac{1}{\sin^2\frac{\pi (L+M)}{n}}\\
&\\
 L=\pm M & \frac{ 4 n^2-1}{3n}-\frac{1}{4n}\frac{1}{\sin^2\frac{\pi L}{n}}-\frac{1}{4n}\frac{1}{\cos^2\frac{\pi L}{n}}\\
&\\
 L= \pm M=\pm\frac{n}{2} & \frac{4 n^2+4}{3n}\\
\end{array}
\right.
\ee
Thus we see that this chain is not orthogonal at tree level,  much as the chain built from $R$ currents. The
basic objects in the theory are in a sense the fermions and these objects are ``composite''s as far as the quantum numbers go.

\section{The spin chain at one loop}\label{1loopsec}

In this section we discuss the structure of the ``one loop''computation of anomalous dimensions
of states with impurities. First, we discuss the symmetry algebra of the theory and the structure of the interaction terms.
Then, we discuss the map to the covering surface and the function $t_L(t)$ at one loop.
Next, we perform the one loop computation of the vanishing anomalous dimension of the vacuum. Finally,
we comment on the one loop computation of states with impurities.

\subsection{The ${\mathcal N}=(4,4)$ supersymmetry algebra and the interaction term}\label{susy}
Let us first  discuss the supersymmetry generators.\footnote{For more details see {\it e.g.}~\cite{David:2002wn,Dabholkar:2007ey}.}
The left-moving supercharges are given by
\be
G^a&=&\sqrt{2}
\left[\begin{array}{c}i\psi^1_I\\-\psi^{2\dagger}_I\\\end{array}\right]
\d X^{1\dagger}_I+\sqrt{2}\left[\begin{array}{c}i\psi^2_I\\\psi^{1\dagger}_I\\\end{array}\right]
\d X^{2\dagger}_I,\\
\hat G^a&=&\sqrt{2}
\left[\begin{array}{c}i\psi^{1\dagger}_I\\\psi^{2}_I\\\end{array}\right]
\d X^{1}_I+\sqrt{2}\left[\begin{array}{c}i\psi^{2\dagger}_I\\-\psi^{1}_I\\\end{array}\right]
\d X^{2}_I.\nonumber
\ee In the above expressions a summation over the copy index $I$ is implied. In the bosonized language this becomes
\be
G^a&=&\sqrt{2}
\left[\begin{array}{c}ie^{i\phi^1_I}\\-e^{-i\phi^{2}_I}\\\end{array}\right]
\d X^{1\dagger}_I+\sqrt{2}\left[\begin{array}{c}ie^{i\phi^2_I}\\ e^{-i\phi^{1}_I}\\\end{array}\right]
\d X^{2\dagger}_I,\\
\hat G^a&=&\sqrt{2}
\left[\begin{array}{c}i e^{-i\phi^{1}_I}\\ e^{i\phi^{2}_I}\\\end{array}\right]
\d X^{1}_I+\sqrt{2}\left[\begin{array}{c}ie^{-i\phi^{2}_I}\\-e^{i\phi^{1}_I}\\\end{array}\right]
\d X^{2}_I.\nonumber
\ee
 The global symmetry of the theory is $SU(2)_R\times SU(2)_I$, where $SU(2)_I$ acts on the generators
in the following way
\be
&&[J_I^i,J_I^j]=i\e_{ijk}J_I^k ,\qquad {\mathcal G}=\left(G^1,\,{\hat G}^2\right),\qquad {\mathcal G}^\dagger=
\left(
\begin{array}{c}
{\hat G}^1\\ G^2\\
\end{array}
\right)\\
&&[J_I^i,{\mathcal G}^a]=\half {\mathcal G}^b\,\s^i_{ba},\qquad [J_I^i,{\mathcal G}^a]=-\half\,\s^i_{ab} {\mathcal G}^{b\dagger}.\nonumber
\ee From here we see that under the $J_3$ charge $G^1$ and $G^2$ have charge $+\half$ while ${\hat G}^1$ and ${\hat G}^2$ have charge $-\half$.
The $SU(2)_R$ acts in the following way,
\be
[J_R^i,J_R^j]=i\e^{ijk} J_R^k,\qquad [J_R^i,\,G^a]=-\half {(\s^i)^a}_b\,G^b,\qquad
[J_R^i,\,{\hat G}_a]=-\half {\hat G}_b{(\s^i)^{b}}_a .
\ee From here we get that generators $G^1$ and ${\hat G}^2$ have charge $+\half$ while
$G^2$ and ${\hat G}^1$ have charge $-\half$. Thus we can write the following
\be
G^1=G^{++},\qquad G^2=G^{-+},\qquad  {\hat G}^1=G^{--},\qquad  {\hat G^2}=G^{+-},
\ee where the first sign is the $SU(2)_R$ charge and the second is the $SU(2)_I$ charge.
We can repeat the same procedure for the anti-holomorphic fields and denote the antiholomorphic
super-charges as $\tilde G^{ab}$, $a=\pm$ and $b=\pm$.

\

We are ready now to discuss the deformation away from the orbifold point.
In a unitary $N=4$ theory, all marginal deformations that preserve the $N=4$ symmetry are obtained from 
chiral fields with dimension $\Delta=1/2$~\cite{Berkovits:1994vy}, such as our twist-two fields $O_2$.
The explicit form of the deformation we will use is
\be
I_{int}&=&\int d^2u\; {\mathcal O}_2+c.c.,\label{inter3}\\
{\mathcal O}_2&=&\lambda \left[G^{--}_{-1/2}\tilde G^{-+}_{-1/2}-G^{-+}_{-1/2}\tilde G^{--}_{-1/2}\right] \, O_2\,.\label{inter2}
\ee  
This deformation has charge zero under both $SU(2)_I$ and $SU(2)_R$. Each one of the two terms above can be turned on with separate coupling with the price of
breaking the global $SU(2)_I$ symmetry.

\noindent We will perform one loop computations on the covering surface using stress-energy tensor method.
 Let us thus lift the deformation explicitly to the covering surface. The explicit computation of the first term in \eqref{inter2} reads
\be\label{firstterm}
&G^{--}_{-1/2}\tilde G^{-+}_{-1/2} \, O_2(u,\bar{u})=&
\\
&\oint\frac{dz}{2\pi i}\oint\frac{d\bar z'}{2\pi i}\left(\sqrt{2}ie^{-i\phi^1_I}\,\d X^1_I+\sqrt{2}ie^{-i\phi^{2}_I}\d X^2_I\right)(z)
\left(-\sqrt{2}e^{-i\bar\phi^{2}_J}\,\bar\d X^{1\dagger}_J+\sqrt{2}e^{-i\bar\phi^{1}_J}\bar\d X^{2\dagger}_J\right)(\bar z')\times&\nonumber
\\
&\times\frac{1}{\sqrt{2N!(N-2)!}}\sum_{h\in S_N}\; e^{ i \frac{1}{4} \sum_{I'=h\cdot 1,\, h\cdot 2} (\phi_{I'}^1 +  \phi_{I'}^2  + \bar{\phi}_{I'}^1 + \bar{\phi}_{I'}^2 )}
   {\sigma}_{h^{-1}\,(1\,2)\,h}(u,\bar u)&\nonumber
\ee 
On the covering surface up to overall constants the integrand of the above expression becomes
\be\label{covfirst}
\frac{2i\left(\frac{\d z}{\d t}\right)^{-\frac{3}{2}}
\;\left(\frac{\d \bar z'}{\d \bar t'}\right)^{-\frac{3}{2}}}{(x-t)^{\half}(\bar x-\bar t')^{\half}}\,
&\biggl[
 :e^{ -i\phi^2(t)-i\bar\phi^1(\bar t')+\frac{i}{2}  (\phi^1(x) +  \phi^2(x)  + \bar{\phi}^1(\bar x) + \bar{\phi}^2(\bar x) )}:
\d X^2\bar\d X^{2\dagger}-&\nonumber\\
 &-:e^{ -i\phi^1(t)-i\bar\phi^2(\bar t')+\frac{i}{2}  (\phi^1(x) +  \phi^2(x)  + \bar{\phi}^1(\bar x) + \bar{\phi}^2(\bar x) )}:
\d X^1\bar\d X^{1\dagger}+&\nonumber\\
&+:e^{ -i\phi^1(t)-i\bar\phi^1(\bar t')+\frac{i}{2}  (\phi^1(x) +  \phi^2(x)  + \bar{\phi}^1(\bar x) + \bar{\phi}^2(\bar x) )}:
\d X^1\bar\d X^{2\dagger}-&\nonumber\\
&-:e^{ -i\phi^2(t)-i\bar\phi^2(\bar t')+\frac{i}{2}  (\phi^1(x) +  \phi^2(x)  + \bar{\phi}^1(\bar x) + \bar{\phi}^2(\bar x) )}:
\d X^2\bar\d X^{1\dagger}\biggr] &
\ee where we assumed that the insertion at $u$ on the base sphere is mapped to $t=x$ on the covering.
We also dropped overall $x$ dependent factors coming from conformal transformations of the operators
as these will cancel out in the stress-energy method of computing correlators. We will drop these kind of terms
everywhere in what follows.
 Near twist two field the map has the property
\be
z(t)\sim v(x) + \half\,a(x)\,(t-x)^2,
\ee
and thus the contour integrals in \eqref{firstterm} can be lifted to the covering surface as
\be
 \oint\frac{dz}{2\pi i}\oint\frac{d\bar z'}{2\pi i}\frac{1}{(x-t)^{\half}(\bar x-\bar t')^{\half}}\;\biggl(\frac{\d z}{\d t}\biggr)^{-\frac{3}{2}}
\;\biggl(\frac{\d \bar z'}{\d \bar t'}\biggr)^{-\frac{3}{2}}
\to \oint\frac{dt}{2\pi i}\oint\frac{d\bar t'}{2\pi i}\frac{1}{(x-t)^{}(\bar x-\bar t')^{}}\, .\nonumber\\
\ee
 Using Cauchy theorem we obtain
\be
&G^{--}_{-1/2}\tilde G^{-+}_{-1/2} \, O_2\sim&\\
&2i\biggl[
e^{ \frac{i}{2}  (\phi^1 -  \phi^2  - \bar{\phi}^1 + \bar{\phi}^2 )}
\d X^2\bar\d X^{2\dagger}
-e^{ \frac{i}{2}  (-\phi^1 +  \phi^2  + \bar{\phi}^1 - \bar{\phi}^2 )}
\d X^1\bar\d X^{1\dagger}+&\nonumber\\
&+e^{ \frac{i}{2}  (-\phi^1 +  \phi^2  - \bar{\phi}^1 + \bar{\phi}^2 )}
\d X^1\bar\d X^{2\dagger}
-e^{ \frac{i}{2}  (\phi^1 -  \phi^2  + \bar{\phi}^1 - \bar{\phi}^2 )}
\d X^2\bar\d X^{1\dagger}
\biggr]\;(x,\bar x)&\, .\nonumber
\ee
Performing the same computation for all the terms is \eqref{inter3}  we finally get
\be
{\mathcal O}^\dagger+{\mathcal O}&\sim& -4i\Re[\lambda]\biggl(e^{- \frac{i}{2}  (\phi^1 -  \phi^2  - \bar{\phi}^1 + \bar{\phi}^2 )}
\d X^{2\dagger}\bar\d X^{2}-e^{- \frac{i}{2}  (\bar \phi^1 -  \bar \phi^2  - {\phi}^1 + {\phi}^2 )}
\bar\d X^{2\dagger}\d X^{2}-\nonumber\\
&&-e^{- \frac{i}{2}  (-\phi^1 +  \phi^2  + \bar{\phi}^1 - \bar{\phi}^2 )}
\d X^{1\dagger}\bar\d X^{1}+e^{ -\frac{i}{2}  (-\bar\phi^1 +  \bar\phi^2  + {\phi}^1 - {\phi}^2 )}
\bar\d X^{1\dagger}\d X^{1}\biggr)+\nonumber\\
&& +4\Im[\lambda]\biggl(e^{- \frac{i}{2}  (-\phi^1 +  \phi^2  - \bar{\phi}^1 + \bar{\phi}^2 )}
\d X^{1\dagger}\bar\d X^{2}+e^{ -\frac{i}{2}  (\bar\phi^1 -  \bar\phi^2  + {\phi}^1 - {\phi}^2 )}
\bar \d X^{2\dagger}\d X^{1}-\nonumber\\
&&-e^{ -\frac{i}{2}  (\phi^1 -  \phi^2  + \bar{\phi}^1 - \bar{\phi}^2 )}
\d X^{2\dagger}\bar\d X^{1}-e^{- \frac{i}{2}  (-\bar\phi^1 +  \bar\phi^2  - {\phi}^1 + {\phi}^2 )}
\bar\d X^{1\dagger}\d X^{2}\biggr)\;.\nonumber\\
\ee
By computing $OPE$s of the above interactions one can verify that there are no dimension $(1,1)$ contact terms.
The conformal dimension on the covering surface of the above operator is $(\frac{5}{4},\frac{5}{4})$ and one obtains
dimension $(1,1)$ on the base by using the fact that for the above operator
$\Delta_{base}=\Delta_n+\frac{\Delta_{cover}}{n}$, where $n$ is the size of the twist field and $\Delta_n$ is the dimension of the bare twist field.
\footnote{Note that the interaction has a very simple ``local'' form on the covering surface. However, this does not imply that
on the base surface the interaction has a simple form.}

\subsection{The map to the covering surface in presence of the interactions}\label{mapsec}

To determine the two point functions of states with impurities in presence of the interactions we
have to lift the  computation to the covering surface and specify the position of the impurities using the function $t_L(t)$.
After turning on the interaction term \eqref{inter3}  in principle, a dressed $n$-cycle twist field mixes
 with dressed $n-1$ cycle twist
 field already in order $\lambda^1$. However, the mixing of two dressed $n$ cycle fields occurs only at even orders in $|\lambda|$.
We will restrict our explicit discussion to the latter case, {\it i.e.} we will study the mixing matrix to the first non trivial order in $\lambda$
between chains of same length. The reason for this is that we will discuss only computations with fermionic impurities,
 and the former mixing is absent for these
as the $X$ correlator becomes a one point function on the covering surface and thus vanishes.
 Using the same techniques as  will be used below one can also perform  calculations of two point functions
of chains of different lengths.  
In what follows we first discuss in detail the map to the covering surface in presence of two twist-two interactions, the first 
non-trivial order contributing to mixing of two chains of same length.
Then we briefly comment on generalizations to maps with more interaction terms. Finally, in the next section we discuss the properties of~$t_L(t)$.

\begin{figure}[tbp]
\begin{center}
$\begin{array}{c}
\\[1.4cm]
\epsfig{file=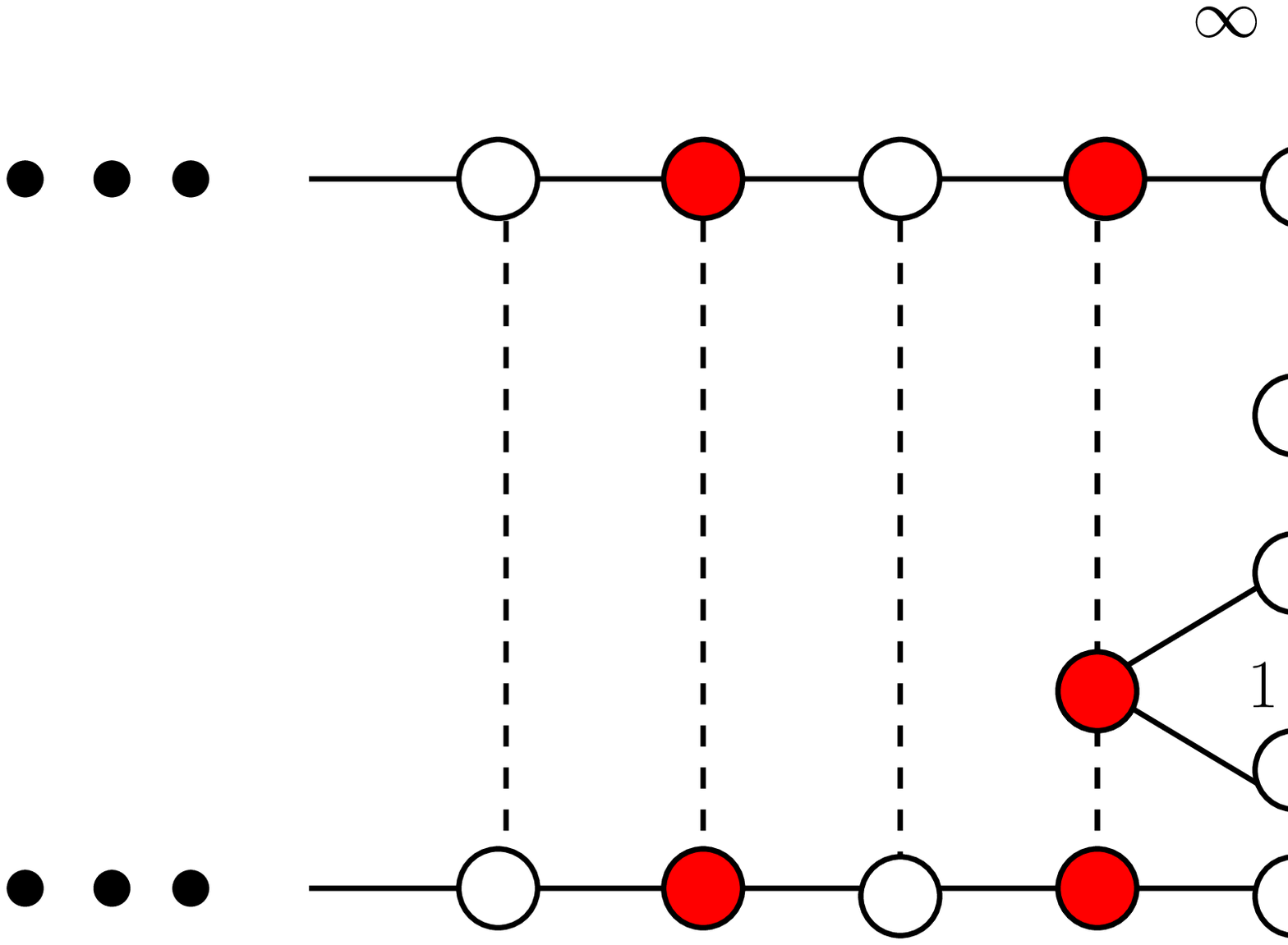,scale=0.2} \\[2.4cm]
\epsfig{file=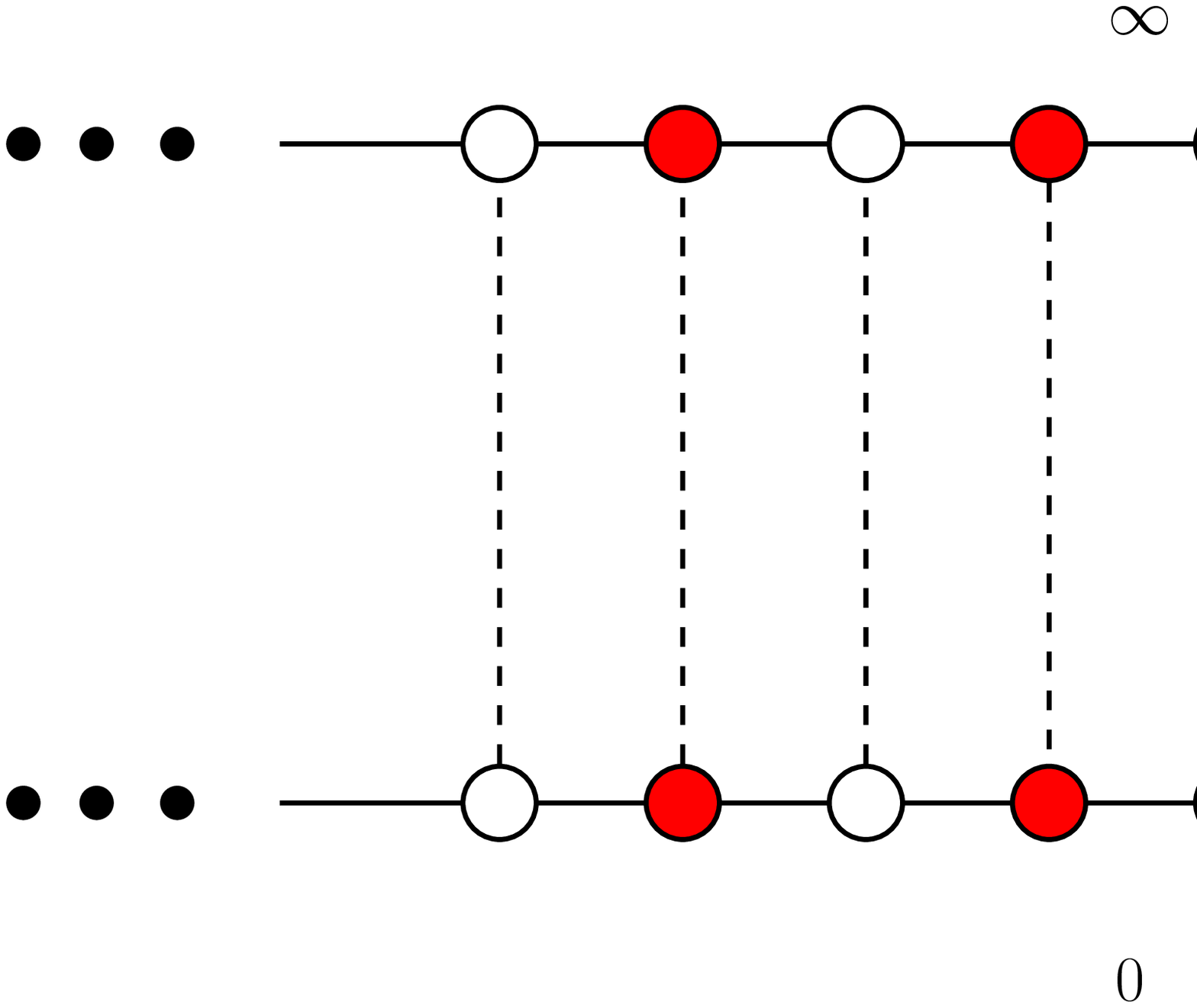,scale=0.2} \\[2.4cm]
\epsfig{file=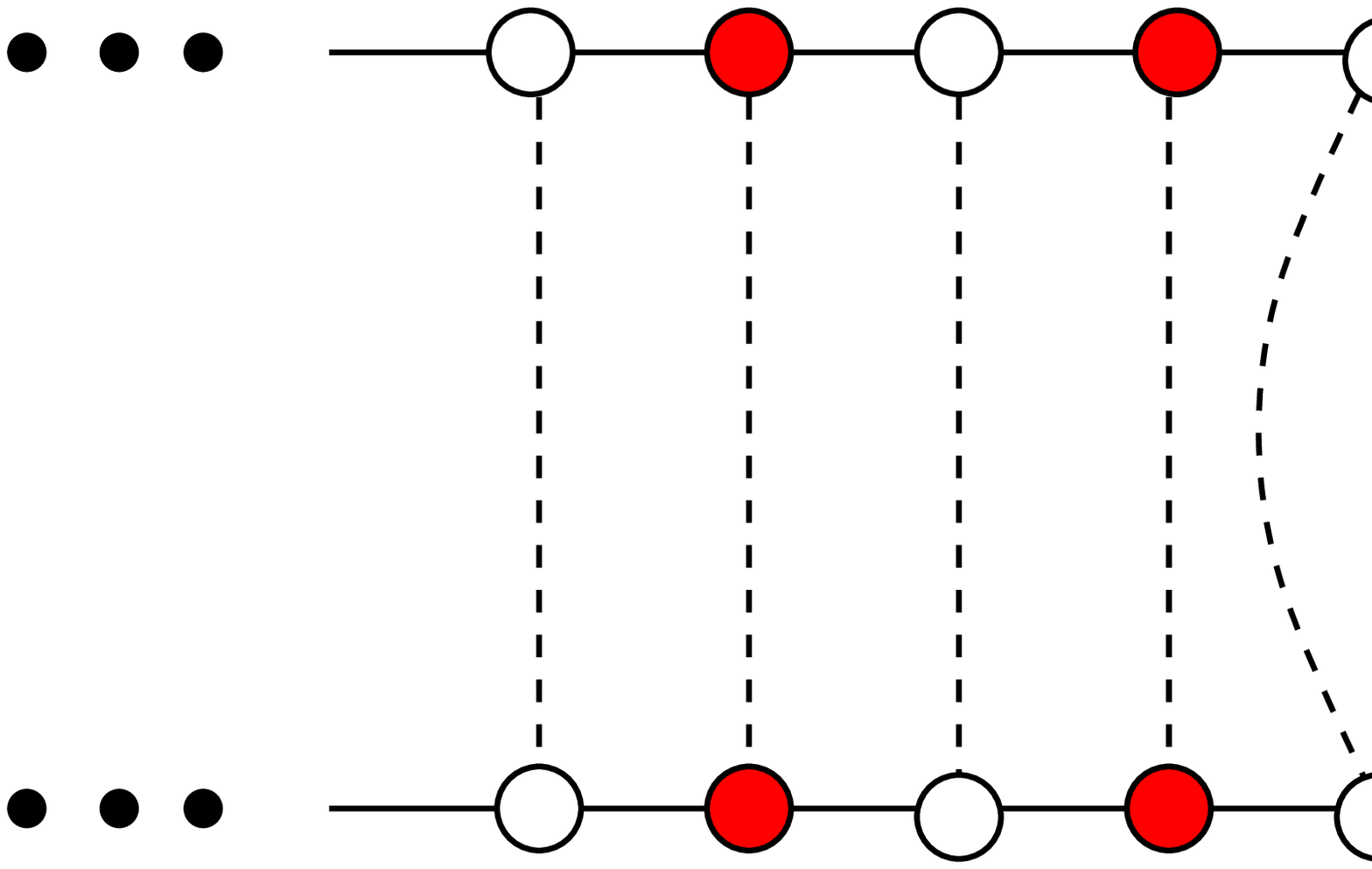,scale=0.2} \\[2.4cm]
\epsfig{file=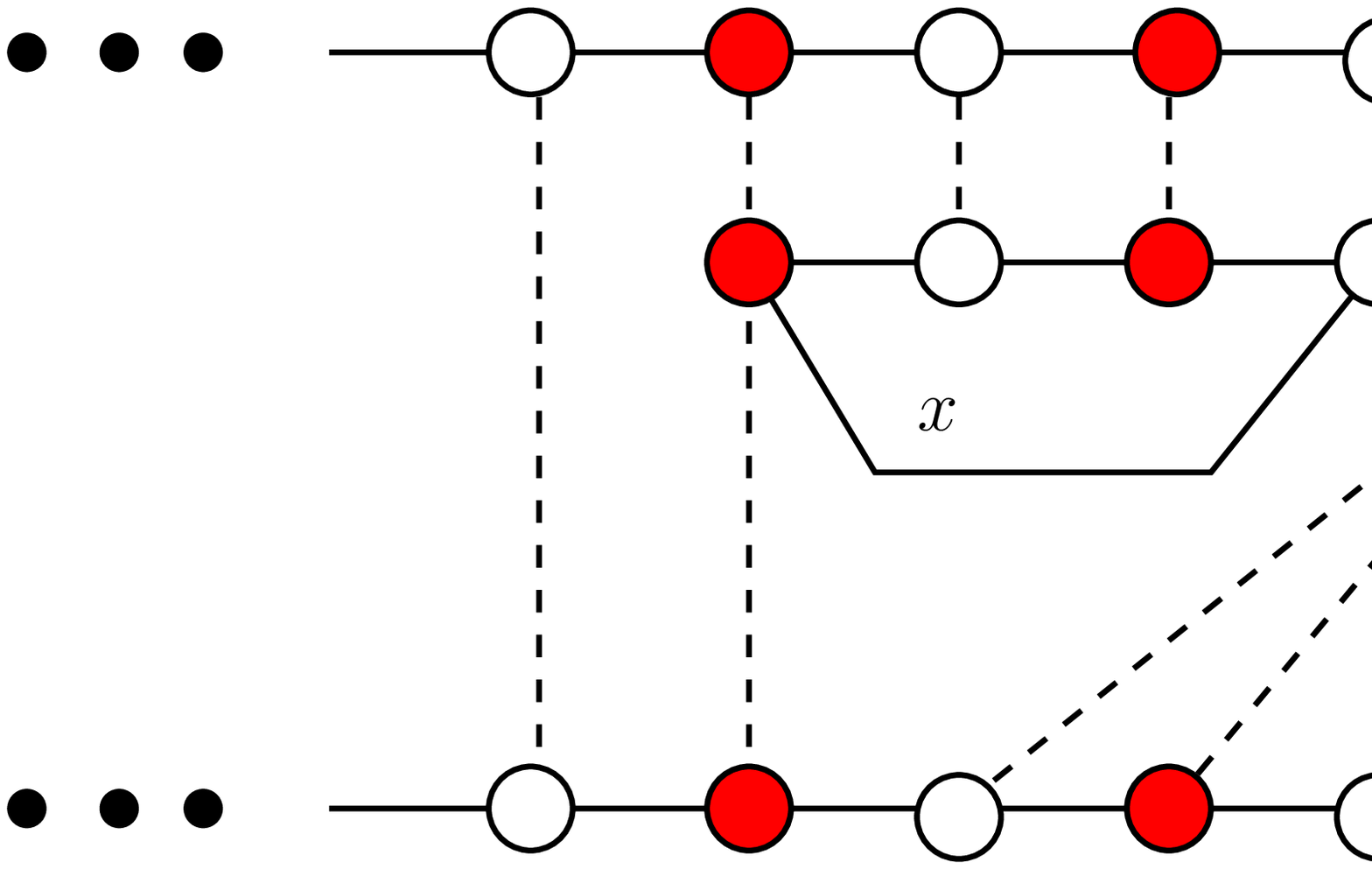,scale=0.2}\\
\\ [0.2cm]
\end{array}$
\end{center}
 \begin{center}
\caption{The diagrams of map $b$ in the vicinity of $x=0$. The bottom diagram is the generic one, there are $n-2$ of these.
 } \label{diags1}
\end{center}
\end{figure}

\begin{figure}[tbp]
\begin{center}
$\begin{array}{c}
\\[3.4cm]
\epsfig{file=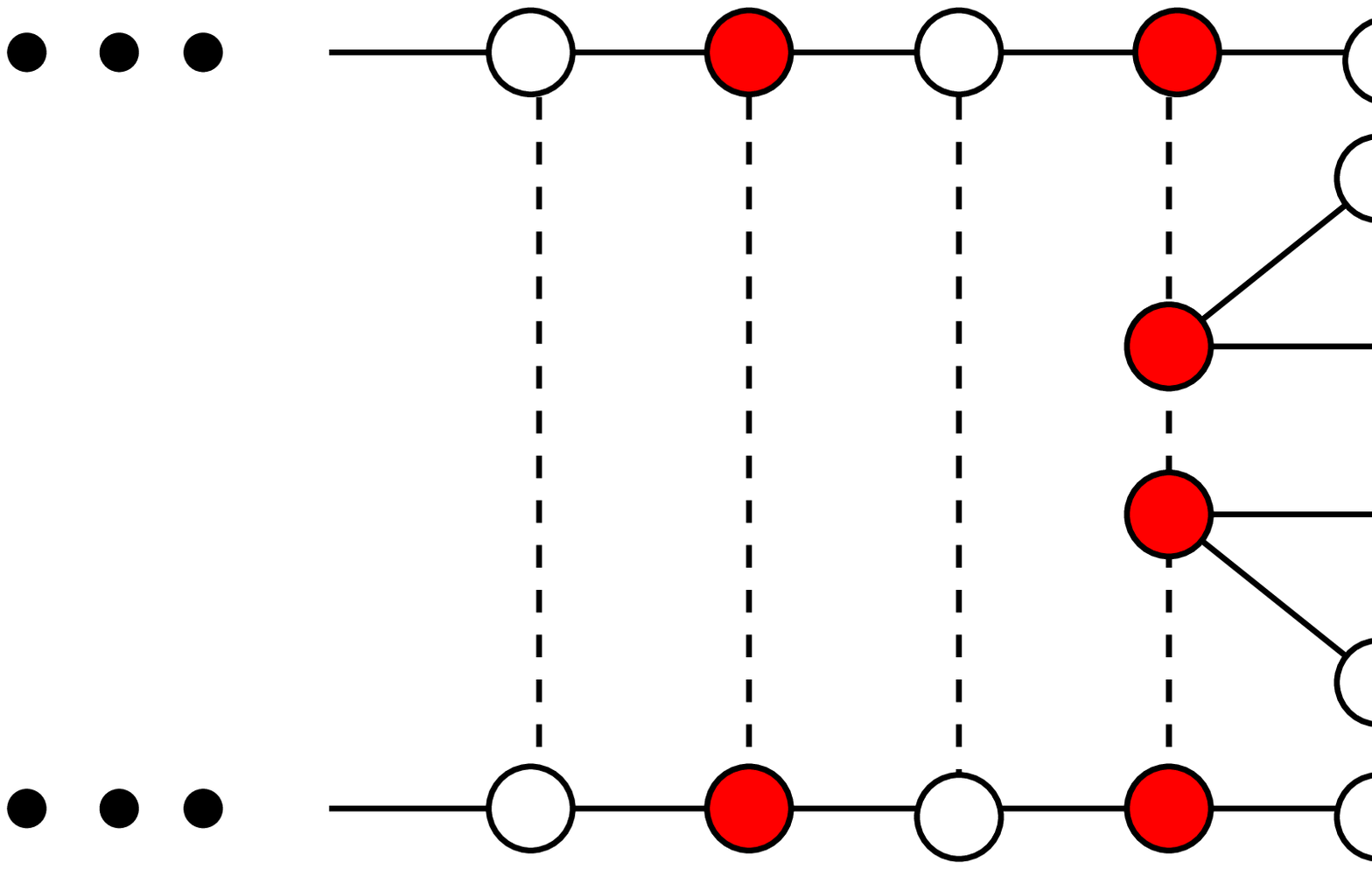,scale=0.2}\\[2.4cm]
\epsfig{file=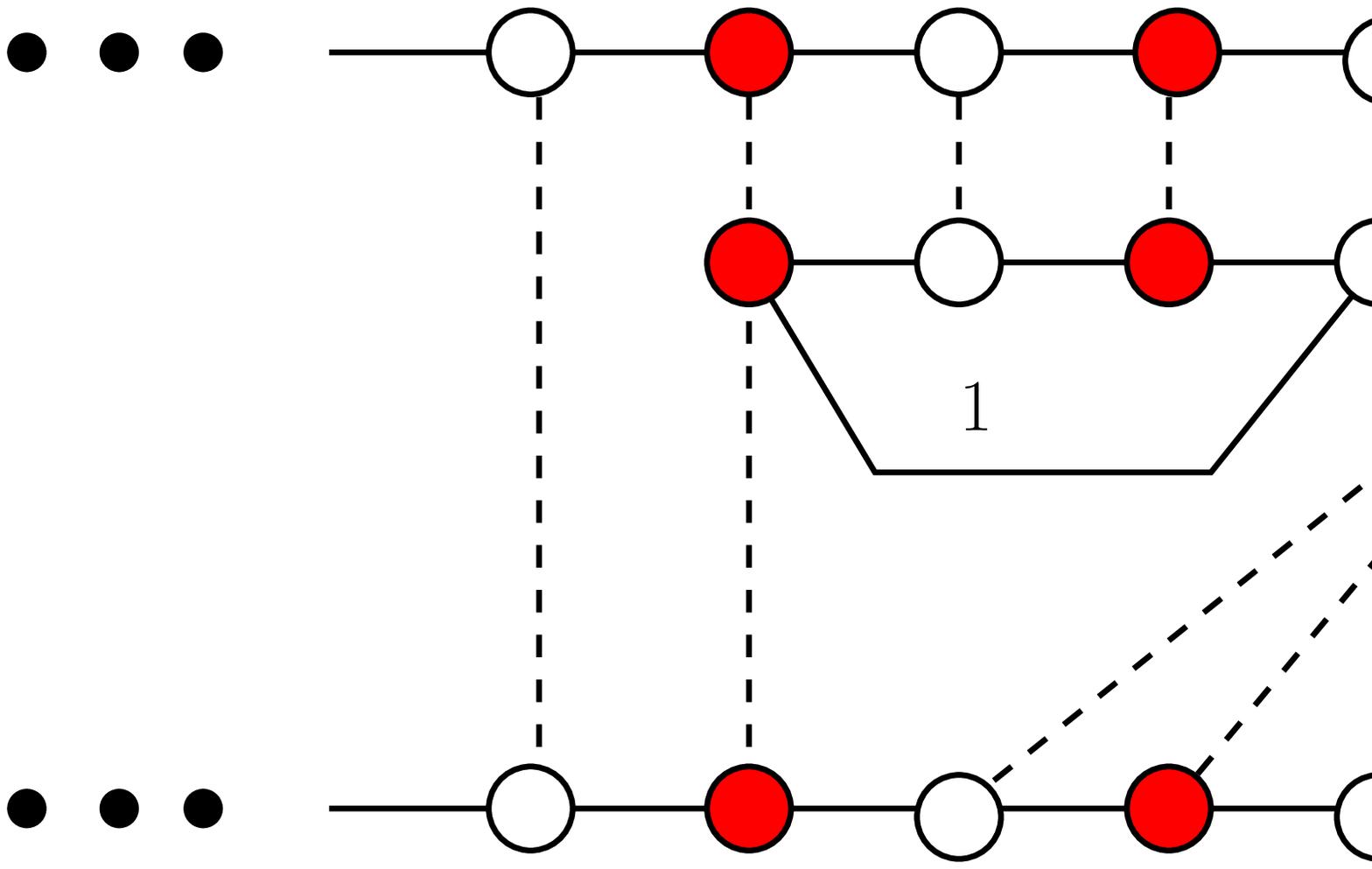,scale=0.2}\\
\\ [0.2cm]
\end{array}$
\end{center}
 \begin{center}
\caption{The diagrams of map $a$ in the vicinity of $x=0$. The bottom diagram is the generic one, there are $n-2$ of these.
 } \label{diags2}
\end{center}
\end{figure}

\begin{figure}[htbp]
\begin{center}
$\begin{array}{c@{\hspace{0.35in}}c}
\epsfig{file=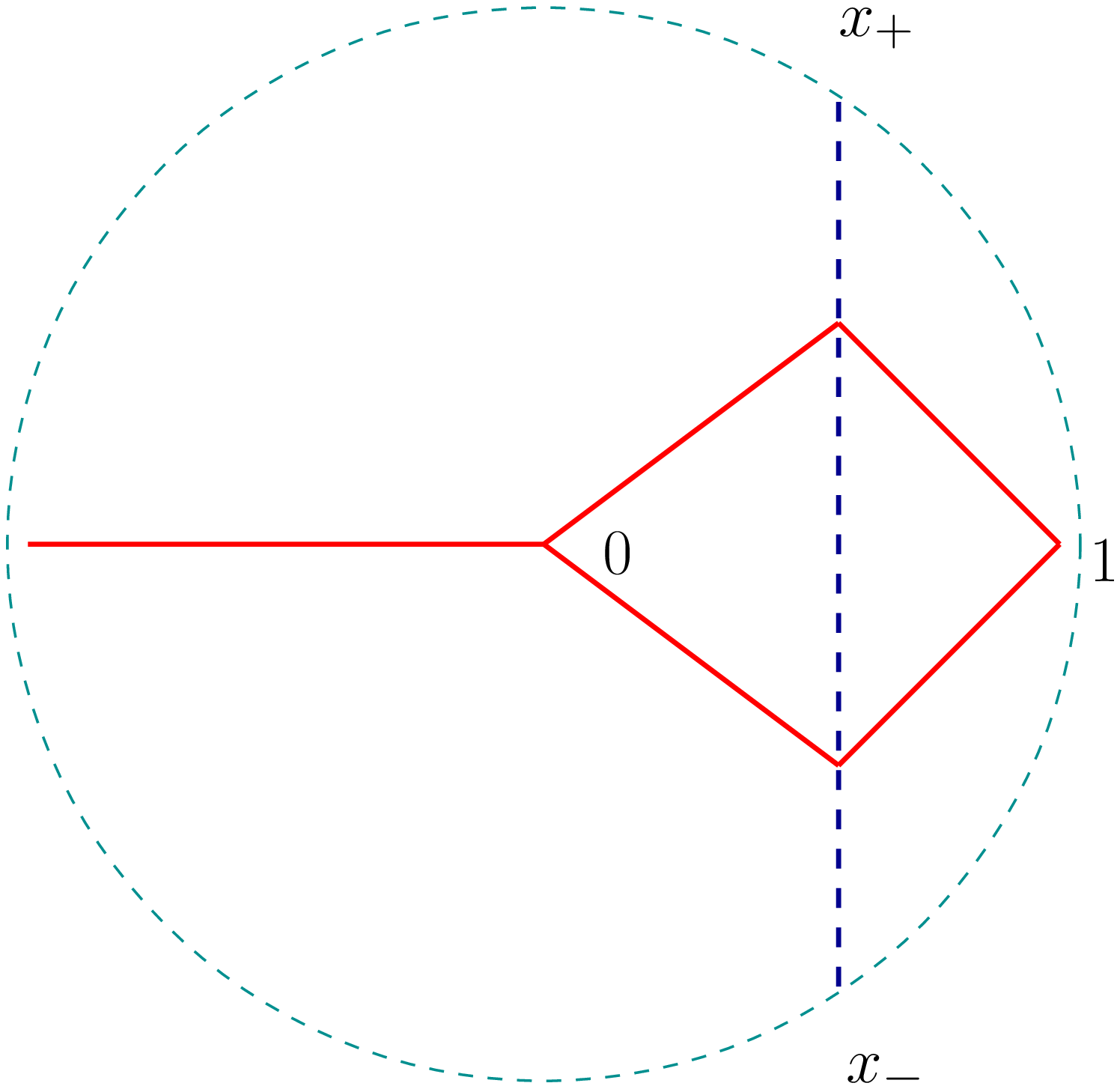,scale=0.3} &
\epsfig{file=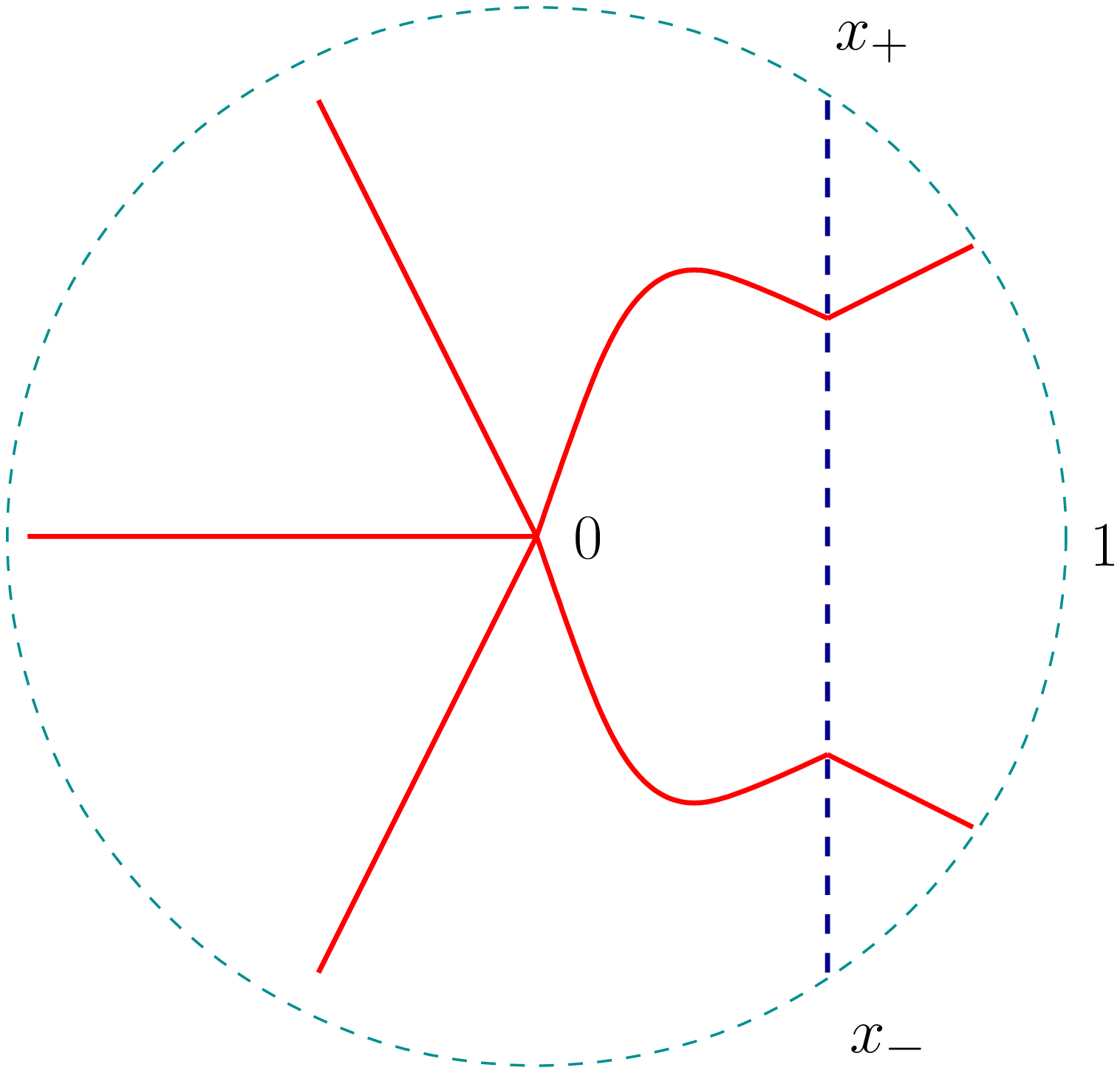,scale=0.3}\\
\\ [0.2cm]
\end{array}$
\end{center}
 \begin{center}
\caption{The schematic structure of the moduli space for $x<1$. On the left we have map $a$ and on the right map $b$. In
this figure $n=4$. Each region of the moduli space corresponds to a diagram. The dashed blue line is the branch cut connecting the two copies
of the moduli space.
 } \label{modulifig}
\end{center}
\end{figure}

Let us specify the map to the covering surface of two twist $n$ fields in presence of two twist two interactions.
The map we construct has twist $n$ fields at $0$ and $\infty$ on the base and the covering.
One twist two field is at $z=1$ on the base and $t=1$ on the covering and the other one
is at $z=u$ on the base and $t=x$ on the cover.
The relevant map  is given by~\cite{Pakman:2009zz}
\be\label{map1}
z(t; x)=\left(\frac{f_2(1)}{f_1(1)}\right)\,\frac{f_1(t)}{f_2(t)} \,,
\ee
where
\be\label{map2}
f_1(t)&=&\,t^n\,\left(1-\frac{n(1+x) \mp \sqrt{n^2 (1-x)^2 + 4 x}}{2x(n+1)}\,t\right) \,,\\
f_2(t)&=&1-\frac{n(1+x) \pm \sqrt{n^2 (1-x)^2 + 4 x}}{2x(n-1)}\,t \,.\nonumber
\ee
We have two choices of the sign before the square root. The map with the $-$ in $f_1$ and $+$ in $f_2$ will
be called  map $a$, and the map with the $+$ in $f_1$ and $-$ in $f_2$ will
be called map $b$ in what follows. We will also write \eqref{map1} as
\be\label{map0}
z(t;x)=C\, t^n\frac{t-t_0}{t-t_\infty}.
\ee For map $a$ we have
\be
t_0&=&\frac{2 (n+1) x}{n(1 +  x) - \sqrt{n^2 ( x-1)^2 + 4 x}}\sim x\,(1+\frac{1}{n})+O(1/n^2),\\
t_\infty&=&\frac{2 (n-1) x}{n(1 +  x) + \sqrt{n^2 ( x-1)^2 + 4 x}}\sim 1-\frac{1}{n}+O(1/n^2), \nonumber
\ee and for map $b$
\be
t_0&=&\frac{2 (n+1) x}{n(1 +  x) + \sqrt{n^2 ( x-1)^2 + 4 x}}\sim 1+\frac{1}{n}+O(1/n^2),\\
t_\infty&=&\frac{2 (n-1) x}{n(1 +  x) - \sqrt{n^2 ( x-1)^2 + 4 x}}\sim x\,(1-\frac{1}{n})+O(1/n^2). \nonumber
\ee Additional useful identities are
\be
\frac{1}{1-t_\infty}-\frac{1}{1-t_0}=n,\qquad t_0\,t_\infty =x.
\ee
  The parameter $x$ is fixed by demanding that $z(x;x)=u$, {\it i.e.}
\be\label{uofx}
u=
\half x^{n-1} \left(2 x + n^2 (x-1)^2  - n(x-1) \sqrt{n^2 (1-x)^2 + 4 x}\right) \,,
\ee which can be more conveniently written as
\be\label{condu}
u^2 + x^{2 n} - u x^{n-1} \left[n^2 (x-1)^2 + 2 x\right]=0.
\ee
Given $u$ we have $2n$ solutions to this equation. Using the diagrammatic language of \cite{Pakman:2009zz} these can be represented
as $2n$ different diagrams.
The diagrams split into two groups, with $n+1$ and $n-1$ diagrams in each group. The groups differ
by the behavior of $x(u)$ in the OPE limits $u=0$ and $u=\infty$. For the first group we have
\be
u\to\infty \qquad :\qquad  u\sim n^2\,x^{n+1},\qquad u\to0 \qquad :\qquad  u\sim \frac{1}{n^2}\,x^{n+1},
\ee and for the second group
\be
u\to\infty \qquad :\qquad  u\sim \frac{1}{n^2}\,x^{n-1},\qquad
u\to0 \qquad :\qquad  u\sim n^2\,x^{n-1}.
\ee Moreover, in the OPE limit $u=1$ we again have different behaviors for the two groups of solutions,
\be\label{OPE1}
 u-1\sim2n\,(x-1)\,,\qquad u-1\sim\frac{n(n^2-1)}{24}(x-1)^3\,,
\ee for the first and the second groups respectively. The diagrams of the first group appear in figure \ref{diags1} and the
 diagrams of the second group appear in figure \ref{diags2}.  There
are three diagrams with a non trivial OPE limit near $u=1$ for the second group and one diagram
with non trivial behavior for the first one, as is clear  from \eqref{OPE1}.
In $u\to 1$ limit the non trivial OPE manifests itself in the diagrams having a shared color between the two
interactions.
The space of possible values of parameter $x$, the ``moduli space'' space of maps~\cite{Pakman:2009zz}, consists of two copies of a sphere glued along a branch cut between
\be
x_+=1-\frac{2}{n^2}\left(1-\sqrt{1-n^2}\right),\qquad x_-=1-\frac{2}{n^2}\left(1+\sqrt{1-n^2}\right),
\ee as can be seen from \eqref{uofx}. Note that in the large $n$ limit $x_+$ and $x_-$ are both very near $x=1$, and effectively the two spheres pinch away.\footnote{
Note that $|x_+|=|x_-|=1$.}
Note also that one moves between maps $a$  and $b$ by crossing with $x$ the branch cut between $x_+$ and $x_-$ and thus
map $a$ and map $b$ correspond to the two copies of the moduli space. In particular map $a$ roughly corresponds to the group of $n-1$ diagrams
and map $b$ to the group of $n+1$ diagrams.  One way to establish this fact is to compute the four point function, {\it i.e.} two chiral
fields with two interactions, for the two maps and look at the behavior in the OPE limits. The details of this computation can be found in appendix \ref{bosmap}.
 The schematic picture of the moduli space  is depicted in figure~\ref{modulifig}.

It is possible to choose another parametrization of the moduli spaces which combine the two copies of the moduli space
\cite{Arutyunov:1997gt}. We present this parametrization in appendix \ref{AFMap}. However, the parametrization with the two copies appearing in this section
will be used in what follows. Additional details on the map needed  in the following sections are collected
in appendix \ref{detailssec}.

Finally, let us comment on higher loop maps.
Adding more interactions it becomes harder to determine the exact map to the covering surface. However, in the leading $1/n$
orders the map is essentially simple. Let us take $2k$ interaction terms, {\it i.e.} $2k$ twist two fields. The generic map takes the following
form
\be
z(t)=C\,t^n\frac{\prod_{j=1}^k(t-t^j_0)}{\prod_{j=1}^k(t-t^j_\infty)}.
\ee Differentiating this map we obtain that the twist fields are located at the solutions of the following equation
\be
 \prod_{j=1}^k(t-t^j_0)\prod_{j=1}^k(t-t^j_\infty)+\frac{t}{n}\sum_{i=0}^k
\left[\prod_{i\neq j}(t-t_0^j)\prod_{j}(t-t^j_\infty)-
\prod_{i\neq j}(t-t_\infty^j)\prod_{j}(t-t^j_0)\right]=0.\nonumber\\
\ee To the leading $1/n$ order the second term is vanishing and we get that all the maps are given by different assignments
of the positions of twist fields $x_i$ ($i=1\dots 2k$ and $x_0=1$) to $t_\infty^i$ and $t^i_0$. For example,
 in the two interaction case we discussed
above in detail we had two possibilities, $(t_0=1,\,t_\infty=x)$ and $(t_0=x,\,t_\infty=1)$. The subleading behavior is also easy to obtain.

\subsection{Evaluating $t_L(t)$}\label{tlsec}

Let us compute the function  $t_L(t)$ for the covering map of the previous section.
 The inverse (to \eqref{map0}) map near $t=0$ is given by the following expansion
\be\label{tdef}
t=\sum_{k=1}^\infty b_k\, z^{\frac{k}{n}}\,.
\ee Then by definition $t_L(t)$ is given by
\be
t_L=\sum_{k=1}^\infty b_k\, z^{\frac{k}{n}}\; p^k,\qquad p=e^{\frac{2\pi i }{n}\, L}.
\ee We want to understand the properties of this function.
First, again by definition
\be\label{zcond}z(t)=z(t_L(t)),\ee {\it i.e.} the points $t_L(t)$ and $t$ correspond to the same position $z$ on the base sphere
but represent different colors.
 This implies the following equality,
\be\label{eqtL}
 t^n \;\frac{t-t_0}{t-t_\infty}={t_L}^n \;\frac{t_L-t_0}{t_L-t_\infty},
\ee which has $n+1$ solutions, and thus there are only $n+1$ functions
satisfying \eqref{zcond}. One solution is trivial $t_L(t)=t$, but the others are not.
Out of $n$ non-trivial solutions $n-1$ correspond to the different choices of $p$,
{\it i.e.} $p=e^{\frac{2\pi i }{n}\, j}$ with $j=1\dots n-1$. We are interested in the
solution with $j=L$. The one extra solution satisfies $t_L(0)=t_0$.

Using the above one can derive the following useful identities,
\be\label{tlrels}
(t-1)(t-x)&=&(t-t_0)(t-t_\infty)+\frac{t}{n}(t_0-t_\infty),\\
\frac{\d t_L}{\d t} &=&\frac{t_L}{t}\frac{t_L-t_0}{t_L-x}\frac{t_L-t_\infty}{t_L-1}\frac{t-x}{t-t_\infty}\frac{t-1}{t-t_0}.\nonumber
\ee

 The function $t_L(t)$ is a solution to a polynomial equation \eqref{eqtL} and thus it is clear that
$t_L(t)$ might have branch cuts and indeed it does. By definition $t_0$ is an image of $z=0$,
and thus for the non trivial solutions of interest to us we have $t_L(t_0)=0$.
 This implies,
using~\eqref{eqtL}, the following
\be
t_L\sim \left[\frac{t_\infty}{t_0}\frac{t_0^n}{t_0-t_\infty}\right]^{1/n} \, (t-t_0)^{\frac1n}.
\ee In analogous way $t_L(t_\infty)=\infty$, and we have
\be
t_L\sim \left[t_\infty^n(t_\infty-t_0)\right]^{1/n} \, \frac{1}{(t-t_\infty)^{\frac1n}}.
\ee From here we see that $t=t_\infty$ and $t=t_0$ are branch points of order $n$
and are connected by a branch cut. This means that the $n$ non-trivial solution of \eqref{eqtL} are different branches of the same function.

\begin{figure}[htbp]
\begin{center}
$\begin{array}{c}
\epsfig{file=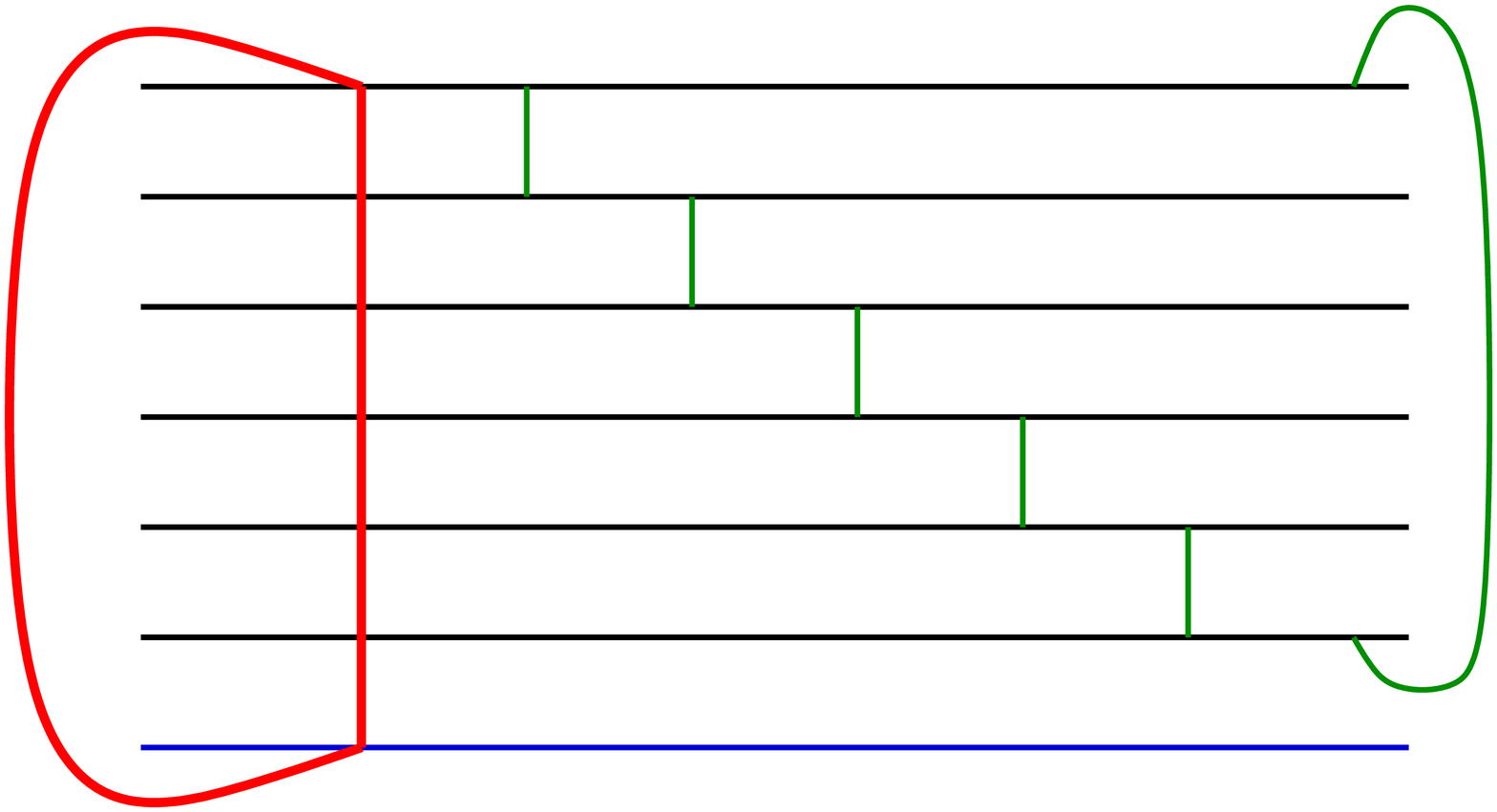,scale=0.35}
\\ [0.2cm]
\end{array}$
\end{center}
 \begin{center}
\caption{ A schematic example of a consistent division of the Riemann surface to sheets.
The different sheets represented by horizontal lines and the cuts by the vertical line.
The special sheet is the bottom one. 
 } \label{cuts}
\end{center}
\end{figure}

Looking for zeros of $\frac{d\, t}{d\, t_L}$ we find that there
are additional branch points of order $2$ at $t$s satisfying  \be\label{2branch}t_L(\hat t_1)=1,\qquad t_L(\hat t_x)=x, \ee
with $\hat t_1\neq 1$ and
$\hat t_x\neq x$.
There are $n-1$ solutions of each type. These solutions are distributed between the different Riemann sheets.
The exact way one distributes these branch point among the sheets depends, by definition, on the choice of the branch cuts.
 One can make the following choice (see figure~\ref{cuts}) .
On each sheet with $t_L(0)=0$ there are exactly two solutions for $\hat t_1$ and two solutions for $\hat t_x$, with two branch cuts connecting
$\hat t_1$ and $\hat t_x$. On a single sheet with $t_L(0)=t_0$ we have a single pair of solutions to \eqref{2branch}. Essentially here
$\left\{\hat t_1=1,\,\hat t_x=x\right\}$
and thus we do not have an additional cut.

 We can think of \eqref{eqtL} as an equation defining a Riemann surface
{\it i.e.} a map between a sphere and another surface. The other surface has two connected
components. The first one is a sphere and corresponds to $t_L(t)=t$. The second one corresponds to a genus $n-1$ surface. The genus
can be calculated through Riemann-Hurwitz formula by noting that, as shown above, we have two ramification points of order $n$,
$2(n-1)$ ramification points of order $2$ and the number of sheets is $n$. The freedom of distributing the ramification points among
the sheets translates to the freedom of dividing the Riemann surface into Riemann sheets.

Let us note the following relations
\be
p\, z^{1/n}&=&t_L\, e^{\frac{1}{n}\ln Q(t_L)},\\ p\, z^{1/n}&=&p\, t\, e^{\frac{1}{n}\ln Q(t)},\nonumber
\ee where we define
\be
z(t)=C\;t^n \;\frac{t-t_0}{t-t_\infty}\equiv t^n\; Q(t).
\ee Using this one can write
\be\label{tlt}
t_L=p\, t \, e^{\frac{1}{n}\ln \frac{Q(t)}{Q(t_L)}}.
\ee
The expression \eqref{tlt} can be consistently expanded in $1/n$ away from the branch cuts, say around $t=0$.
 To leading orders in $1/n$  gives
\be
t_L=p\,t\,\left(1+\frac{1}{n}\,\ln \frac{Q(t)}{Q(p\,t)}+\frac{2\pi i}{n}\,l\dots\right)\,.
\ee  The term proportional to $l$ takes into account possible crossing of a branch cut of the $\ln$, and in vicinity of $t=0$ by definition
$l=0$.
For our specific map $\ln \frac{Q(t)}{Q(p\,t)}$ in the large $n$ limit evaluates to 
\be
 \ln \frac{Q(t)}{Q(p\,t)}=(-)^{\s+1}\ln\left[\frac{(t-x)(p t -1 )}{(p t -x)(t-1)}\right],
\ee where $\s=1$ for map $a$ and $\s=0$ for map $b$.
Note that this result makes sense if $t$ is away from $x$ and $1$, thus we have to be either sufficiently close to $0$ or $\infty$.\footnote{
In large $n$ limit we have
$t_0\sim x^{\s}$ and $t_\infty\sim x^{1-\s}$.}

Let us compute the points  $\hat t_x$ and  $\hat t_1$ in the limit $1\ll L\ll n$.
Near $t_L=1$ we define $t_L=1+\e$ and $t=\chi$. From here we obtain that
\be
1+\e = (1+\frac{2\pi i}{n}L)\chi\,e^{\frac{1}{n}\ln\frac{\chi-t_0}{\chi-t_\infty}\frac{1+\e-t_\infty}{1+\e-t_0}}.
\ee  Remembering the behavior of $t_{0,\infty}$ for the two maps and assuming that $\chi\sim 1+\delta$  we deduce
\be\label{tL1ed}
(a)\quad :\quad && \e n-\ln(1+\e n)=2\pi iL+\delta n-\ln(1+\delta n),\\
(b)\quad :\quad && \e n+\ln(1-\e n)=2\pi iL+\delta n+\ln(1-\delta n).\nonumber
\ee
The points with $\e=0$ are $\hat t_1$ by definition. We get two such points on every sheet and the solutions are
\be
(a)\quad :\quad &&\hat t_1= 1-\frac{2\pi i }{n}L+\frac{1}{n}\ln(-2\pi iL)\\
&&\hat t'_1= 1-\frac{2\pi i }{n}(L+1)+\frac{1}{n}\ln(-2\pi i L)\nonumber\\
(b)\quad :\quad &&\hat t_1= 1-\frac{2\pi i }{n}L-\frac{1}{n}\ln(2\pi iL)\nonumber\\
&&\hat t'_1= 1-\frac{2\pi i }{n}(L-1)-\frac{1}{n}\ln(2\pi i L),\nonumber
\ee where the second solution comes from encircling once the branch point of the $\log$ on the \textit{l.h.s} of \eqref{tL1ed}.
The branch point is at $t_L=t_\infty$ for map $a$ and at $t_L=t_0$ for map $b$. Note that on generic sheets there are no points such that
$t_L(t)=t_{0,\infty}$ and these points are ``swallowed'' by the branch cuts.

Near $t_L=x$ we define $t_L=x(1+\e)$ and $t=\chi$.
We obtain that
\be
x(1+\e) = (1+\frac{2\pi i}{n}L)\chi\,e^{\frac{1}{n}\ln\frac{\chi-t_0}{\chi-t_\infty}\frac{x(1+\e)-t_\infty}{x(1+\e)-t_0}}.
\ee  Assuming that $\chi\sim x(1+\delta)$  one deduces
\be\label{tLxed}
(a)\quad :\quad && \e n+\ln(1-\e n)=2\pi i L+\delta n+\ln(1-\delta n),\\
(b)\quad :\quad && \e n-\ln(1+\e n)=2\pi i L+\delta n-\ln(1+\delta n).\nonumber
\ee
The points with $\e=0$ are $\hat t_x$ by definition. We get two such points on every sheet and the solutions are
\be
(a)\quad :\quad &&\hat t_x= x\left(1-\frac{2\pi i }{n}L-\frac{1}{n}\ln(2\pi iL)\right)\\
&&\hat t'_x= x\left(1-\frac{2\pi i }{n}(L+1)-\frac{1}{n}\ln(2\pi i L)\right)\nonumber\\
(b)\quad :\quad &&\hat t_x= x\left(1-\frac{2\pi i }{n}L+\frac{1}{n}\ln(-2\pi iL)\right)\nonumber\\
&&\hat t'_x= x\left(1-\frac{2\pi i }{n}(L-1)+\frac{1}{n}\ln(-2\pi i L)\right),\nonumber
\ee
\begin{figure}[htbp]
\begin{center}
$\begin{array}{c@{\hspace{0.4in}}c}
\epsfig{file=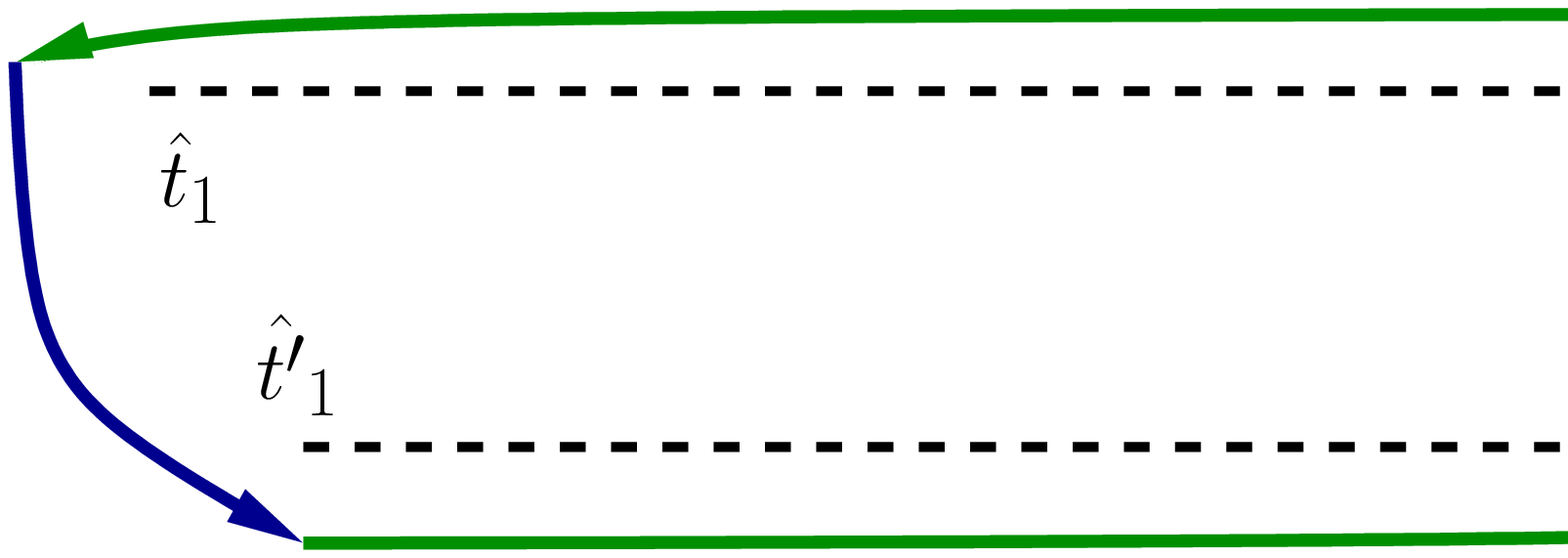,scale=0.35}&\epsfig{file=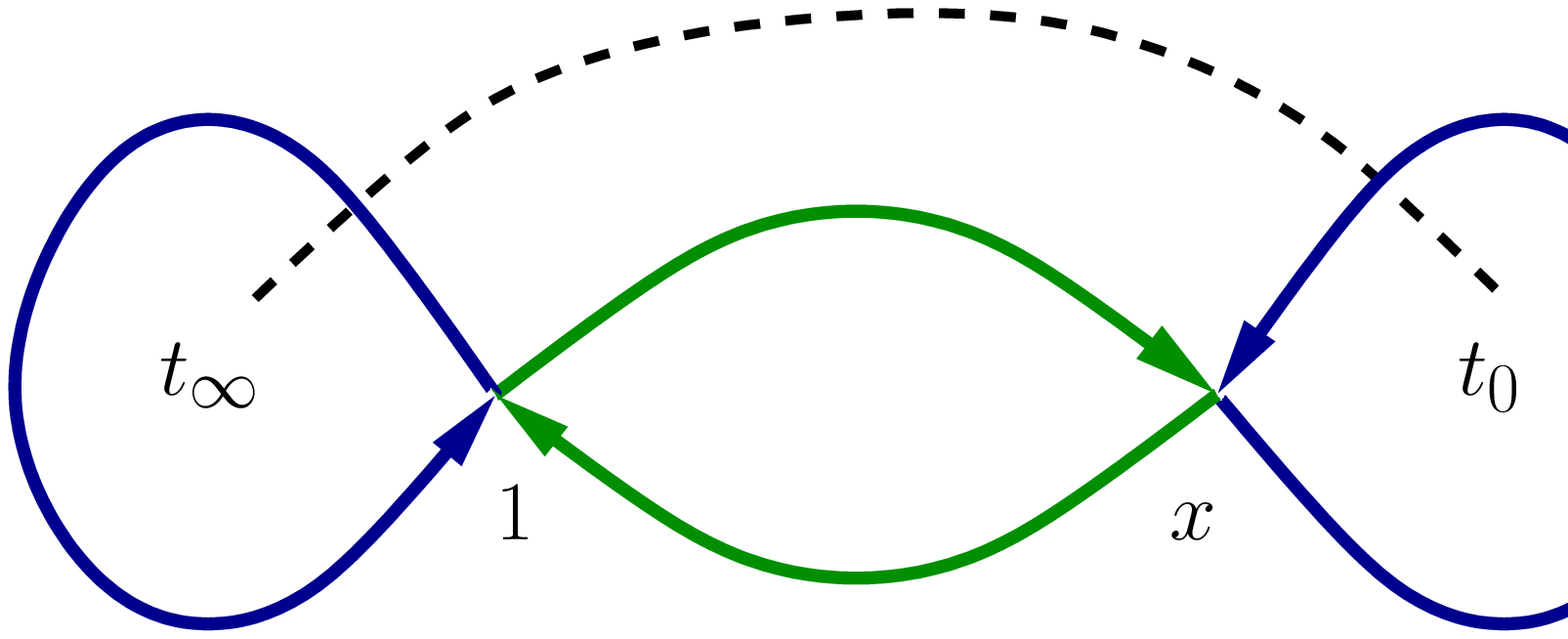,scale=0.35}
\\ [0.2cm]
\end{array}$
\end{center}
 \begin{center}
\caption{ A contour around the two order two cuts on the $t$ sheet transforms to a contour around
the order $n$ cut on the $t_L$ sheet. The two green segments on the right are on different sheets and thus the integral
over the green loop is in principal not vanishing.  This illustration is for map $a$, and for map $b$
$t_0$ has to be interchanged with $t_\infty$.
 } \label{cutt1tx}
\end{center}
\end{figure}
Note also that a loop around the two branch cuts between the images of $t_L=1,x$ under $t_L(t)$ maps to a loop around the cut
between $t_\infty$ and $t_0$. This fact is illustrated in figure \ref{cutt1tx}. Thus, a contour integral of a function of $t$ and $t_L$
around the cut between  $t_\infty$ and $t_0$, integrated over $t$, can be brought to an integral over $t_L$ and around the two additional cuts.
Equivalently, it is equal to an integral around the two additional cuts with $L$ traded for $-L$. This fact will play a role in what follows.

\subsection{Non-renormalization of the chiral vacuum at one loop}\label{vacsec}
As a first step toward discussing the one loop structure of the two point functions of states with impurities
 we will compute first correction to the two point functions of operators without impurities in
the deformed symmetric product CFT. These two point functions are protected~\cite{deBoer:2008ss} and thus the corrections will vanish.

Let us discuss the general structure of the one loop computation.
The first correction to the two-point function is given by
\be
\int d^{2}\!z_2 \, d^2\!z_3  \langle O_1(z_1) {\mathcal O}_2(z_2) {\mathcal O}_3(z_3) O_4(z_4) \rangle,
\label{4pfinteg}
\ee where ${\mathcal O}_2$ and ${\mathcal O}_3$ are the interaction vertices.
Using $\Delta_1 = \Delta_4 =\Delta$ and $\Delta_2=\Delta_3=1$, global conformal invariance fixes the form
\be
\langle O_1(z_1) {\mathcal O}_2(z_2) {\mathcal O}_3(z_3) O_4(z_4) \rangle =
\frac{G(u,\bar{u})}{|z_{24}|^4 |z_{13}|^4 |z_{14}|^{4 \Delta -4}   },
\ee
where
\be
u = \frac{z_{12} z_{34}}{z_{13} z_{24}}.
\ee
We can change now the integration variables in (\ref{4pfinteg}) to $u,z_3$, and using
\be
\left| \frac{\d (u,z_3)}{\d (z_2,z_3)} \right| = \left| \frac{ z_{34} z_{14}}{z_{13} z_{24}^2 } \right| \,,
\ee
eq. (\ref{4pfinteg}) becomes
\be
\frac{1}{|z_{14}|^{4 \Delta}} \int \!\! d^2 \!z_3 \frac{|z_{14}|^2}{|z_{31}|^2 |z_{34}|^2 }
\int \!\!  d^2 \! u \, G(u,\bar{u}) \,.
\ee
The first integral is divergent and must be regulated as
\be
\int \!\! d^2 \! z_3 \frac{|z_{14}|^2}{|z_{31}|^2 |z_{34}|^2 } = \int \frac{d^2 \! w}{|w|^2 |w+1|^2} = 2\pi\log \Lambda\,,
\ee
where $w= z_{31}/z_{14}$, and $\frac{1}{\Lambda}	$ is a cutoff in $|w|$. So the expression (\ref{4pfinteg}) is finally
\be
2\pi\,\frac{\log \Lambda}{|z_{14}|^{4 \Delta}}  \int \!\!  d^2 \! u \, G(u,\bar{u}) \,,
\ee
and we expect the integral over $u, \bar{u}$ to give a finite contribution.

\

We will now explicitly compute $G(u,\bar{u})$.
We do the computation using stress-energy tensor technique.
On the base sphere the two chiral operators are at $z=0,\, \infty$ and the two interactions at $z=1,\, u$. On the
covering sphere the two chiral operators are at $t=0$ and $t=\infty$
and the two interactions are at $t=x$ and $t=1$.
The stress-energy tensor is given by
\be
T(z) &=& -\sum_{i=1}^2\left[\d X^i_I(z) \d X^{i\dagger}_I(z)  +\frac12  \d \phi^i_I(z) \d \phi^i_I(z)\right]
\\
&=& - \lim_{w \rightarrow z}   \sum_{i=1}^2\left[   \d X^i_I(z) \d X^{i\dagger}_I(w)
+  \half\d \phi^i_I(z) \d \phi^i_I(w)    + \frac{6N}{(z-w)^2} \right]\, .
\label{ssing}
\ee The bosonic correlator is given by
\be
&&I_b=\sum_{i=1}^2\frac{\langle \d X^i_I(z) \d X^{i\dagger}_I(w) o_{n}(\infty) {\mathcal I}(1){\mathcal I}(u) o_{n}(0) \rangle_j}
{\langle o_{n}(\infty) {\mathcal I}(1){\mathcal I}(u) o_{n}(0) \rangle_j}=\\
&&\quad\quad= -4\,\frac{t'_{j,I}(z) t'_{j,I}(w)}{(t_{j,I}(z)-t_{j,I}(w))^2}
- 2\;(t'_{j,I}(z))^2\frac{(x-t_1)^2}{(t-1)^2(t-x)^2}  \,.\nonumber
\ee  The fermionic contribution is
\be\label{ferms}
&&I_f=\sum_{i=1}^2\frac{\langle \d \phi^i_I(z) \d \phi^{i}_I(w) o_{n}(\infty) {\mathcal I}(1){\mathcal I}(u) o_{n}(0) \rangle_j}
{\langle o_{n}(\infty) {\mathcal I}(1){\mathcal I}(u) o_{n}(0) \rangle_j}=\\
&&\quad\quad= -2\,\frac{t'_{j,I}(z) t'_{j,I}(w)}{(t_{j,I}(z)-t_{j,I}(w))^2}
- \frac{1}{2}\;(t'_{j,I}(z))^2\biggl[\left(\frac{1}{t-1}-\frac{1}{t-x}\right)^2+\frac{(n-1)^2}{t^2}\biggr]  \,.\nonumber
\ee
Then we define
\be
g_j(z,u) =  \frac{\langle T(z)o_{n}(\infty) {\mathcal I}(1){\mathcal I}(u) o_{n}(0) \rangle_j}
{\langle o_{n}(\infty) {\mathcal I}(1){\mathcal I}(u) o_{n}(0) \rangle_j} \,\,,
\label{gzu}
\ee and obtain
\be\label{gzuf}
g_j(z,u) &=& \frac{6}{12} \sum_{I=1}^{\tilde n} \{t_{j,I},z \}+ \sum_{I=1}^{\tilde n}\frac{(t'_{j,I}(z))^2}{2}\biggl(
2\;\frac{(x-1)^2}{(t-1)^2(t-x)^2}+\\
&&+\frac{1}{2}\biggl[\left(\frac{1}{t-1}-\frac{1}{t-x}\right)^2+\frac{(n-1)^2}{t^2}\biggr]
\biggr) ,\nonumber
\ee
where
$\{t,z \}$
is the Schwartz derivative,
\be
\{t, z\} = \frac{t'''}{t'} -\frac32\left(\frac{t''}{t'}  \right)^2 = \left( \frac{t''}{t'} \right)' -\frac12 \left(\frac{t''}{t'} \right)^2\, ,
\ee The number $\tilde n$ is the number of active colors in the vicinity of the field at $z=u$, {\it i.e.} in our case $\tilde n = 2$. Finally,
remembering that
\be
\d_u \ln G_j(u) =\left\{g_j(z,u)\right\}_{\frac{1}{z-u}} \,,
\label{eqfg}
\ee we obtain the differential equation for $G(u)$
\be\label{diffeq}
\d_u \ln G(x(u))=&&\biggl\{\left[\left(\frac{t''}{t'}\right)'-\half\left(\frac{t''}{t'}\right)^2\right]
+(t')^2\biggl(
2\;\frac{(x-1)^2}{(t-1)^2(t-x)^2}+\\
&&+\frac{1}{2}\biggl[\left(\frac{1}{t-1}-\frac{1}{t-x}\right)^2+\frac{(n-1)^2}{t^2}\biggr]
\biggr)
\biggr\}_{\frac{1}{z-u}}\nonumber.
\ee	
Note also that in this case we can write
\be
v'(x)\d_u \ln G(x(u)) =\d_x \ln G(x(u)),
\ee as the only dependence on $u$ on the right hand side of the differential equation is through $x(u)$.
\
Using the explicit map (see appendix~\ref{detailssec} for details) this equation can be integrated to obtain
\be\label{G0}
&&G^{a}_0(x)= C\;x^{2-n}\,\left(x-1\right)^{-4}\,\,\left(1+x+\sqrt{n^2(x-1)^2+4x}\right)\,\\&&\left(n^2(x-1)-2x-n\sqrt{n^2(x-1)^2+4x}\right)^{-\half}\,
\left(n^2(x-1)+2+n\sqrt{n^2(x-1)^2+4x}\right)^{\frac{3}{2}},\nonumber
\ee
\be
&&G^{b}_0(x)= 4(1-n^2)^2C\;x^{1-n}\,\left(x-1\right)^{-2}\,\,\left(1+x+\sqrt{n^2(x-1)^2+4x}\right)^{-1}\,\\&&\left(n^2(x-1)-2x-n\sqrt{n^2(x-1)^2+4x}\right)^{\half}\,
\left(n^2(x-1)+2+n\sqrt{n^2(x-1)^2+4x}\right)^{-\frac{3}{2}},\nonumber
\ee where $C$ is an overall constant not fixed by our method
of computation. Note that the second expression is equal to the first one after we change the sign in front of the square roots.
Thus, we can only take the first expression and remember that $x$ takes values in a double cover of a sphere.
The OPE behavior on the two covers is different.
Note also the only singularities of the four point function occur when $x=0,1$ or $x=\infty$.

 A simple check of this equation is to take $x\to 1$ in $G^{b}_0(x)$. In this limit $u\to1$ and
corresponds to an OPE limit of the two twist-two interactions sharing both colors
and thus annihilating each other, see equation \eqref{OPE1}.
 The four point function scales as $G(u)\sim \frac{1}{(1-u)^2}$ as expected as the interactions have dimension one.

We can easily integrate \eqref{G0} over the moduli space of maps
\be
&&\sum_{j=1}^{2n}\int d^2 u G(x_j(u),\bar x_j(\bar u))=\int d^2x |v'(x)|^2 G(x)G(\bar x)=
\int d^2x \left|v'(x) G(x)\right|^2=\\
&&=2|n(n^2-1)C|^2\int d^2x \left|\frac{\sqrt{2 x + n (1 + x) \sqrt{n^2 ( x-1)^2 + 4 x} + n^2 (1 + x^2)}}{ (x-1)^2\sqrt{n^2(x-1)^2+4x}}\right|^2=\nonumber\\
&&=\sqrt{2}\,|n(n^2-1)C|^2\int d^2x \left|\frac{n(x+1) + \sqrt{n^2 ( x-1)^2 + 4 x}}{ (x-1)^2\sqrt{n^2(x-1)^2+4x}}\right|^2,\nonumber
\ee where the integration over $x$ presumes integration over the double cover. The sum in the first line is over all the solution
to \eqref{condu}, {\it i.e.} over all the diagrams in figures \ref{diags1} and \ref{diags2}.

We are interested in working in large $n$ limit.\footnote{Note that one can repeat the following
with finite $n$.
Finite $n$ result can be found in appendix \ref{AFMap} using an alternative parametrization of the covering map.} In this limit we obtain the following simple equation
\be \label{finint}
\sum_{j=1}^{2n}\int d^2 u G(x_j(u),\bar x_j(\bar u))\sim|2n^3\,C|^2\int d^2x \left[\left|\frac{x}{ (x-1)^3}\right|^2
+\left	|\frac{1}{ (x-1)^3}\right|^2\right],
\ee where on the \textit{r.h.s} the two terms come from the two different maps. Using the following
\be\label{gamas}
\int d^2z\;z^a{\bar z }^{\bar a}(1-z)^b(1-\bar z)^{\bar b}=\pi\frac{\Gamma(1+a)\Gamma(1+b)\Gamma(-\bar a-\bar b-1)}
{\Gamma(-\bar a)\Gamma(-\bar b)\Gamma(a+b+2)},
\ee and plugging $a=\bar a=1,0$, $b=\bar b=-3$ we get that \eqref{finint} vanishes
and thus the chiral operators do not acquire an anomalous dimension.

 Let us compute the first $1/n$ correction. We have
\be
\delta_1=-\frac{1}{n} \,\delta_0&&+\sqrt{2}\,|n\,C|^2\int d^2x \left|\frac{2x}{ (x-1)^3 }\right|^2
\left(\frac{1}{(x-1)}-\frac{2x}{(x-1)^2}+c.c.\right)-\nonumber\\
&&-\sqrt{2}\,|n\,C|^2\int d^2x \left|\frac{2}{ (x-1)^3 }\right|^2
\left( \frac{x}{(x-1)}+\frac{2x}{(x-1)^2}+c.c.\right),\nonumber
\ee and again using \eqref{gamas} this is vanishing.

We have mentioned that there are no dimension $(1,1)$ contact terms and one can verify this from the above explicit expression
by taking appropriate $OPE$ limits. In appendix \ref{AFMap} we explicitly show  this in a different setup.

\subsection{Comments on one loop with impurities}
We are now ready to add impurities to the computation. In previous sections we have computed
the two point function of operators with impurities but without the interactions, and two points of chiral operators
without impurities but with interaction insertions. In this section we will combine these results to discuss the general structure of the computation
of  anomalous dimensions of the spin chain in one loop. For simplicity we
will discuss only states with two holomorphic fermionic impurities of type $A$, {\it i.e.} one loop correction to 
\eqref{Lnorm}.

We add impurities by ``dressing'' the chiral vacuum with contour integrals. It is convenient to keep the contour integrals and
to compute the six point function of the four fermionic dressings and the two chiral fields first. The dressings are located at
$t,\, t_L$ and $t',\, t'_M$. Note that because the dressing is given in terms of untwisted sector fields this six point function
is simply given by a product of the one loop vacuum result we obtained in previous section and the free field contractions of the dressing fermions
and the fermions appearing in the interaction vertices. This free field computation gives the following result
\be
&&G_1^{(a/b)}(t,t_L,t',t'_M,x)\equiv(t-t_L)(t'-t'_{M})\sqrt{\frac{\d t_L}{\d t}\frac{\d t'_{M}}{\d t'}}
\frac{t^{\half(1-n)}t_L^{\half(1-n)}{t'}^{\half(n-1)}{t'}_{M}^{\half(n-1)}}
{(t-t')(t_L-t')
(t-t'_{M})(t_L-t'_{M})}\times\nonumber\\
&&\qquad\qquad\times \left[\sqrt{\frac{t-1}{t-x}\frac{t_L-1}{t_L-x}\frac{t'-x}{t'-1}\frac{t'_{M}-x}{t'_{M}-1}}+
\sqrt{\frac{t-x}{t-1}\frac{t_L-x}{t_L-1}\frac{t'-1}{t'-x}\frac{t'_{M}-1}{t'_{M}-x}}\right],
\ee where the index $(a/b)$ refers to the map with which we evaluate $t_L$ and $t'_M$.
The terms in the second line come from contractions with the interactions.
Thus, to leading order in $1/n$, the two point function with impurities is given by
\be
\langle M|L\rangle_{1-loop}=|2n^3\,C|^2\int d^2x \;\;\biggl\{ &&
\left|\frac{x}{ (x-1)^3}\right|^2\;\oint\frac{dt}{2\pi i}\; \oint \frac{dt'}{2\pi i}\,G_1^{(a)}(t,t_L,t',t'_M,x)+\nonumber\\
&&+\left|\frac{1}{ (x-1)^3}\right|^2\;\oint\frac{dt}{2\pi i}\; \oint \frac{dt'}{2\pi i}\,G_1^{(b)}(t,t_L,t',t'_M)
\biggr\}.
\nonumber\\
\ee

\

Let us discuss the structure of the above computation. First, we have to 
we have to evaluate the following contour integrals
\be
&& \oint\frac{dt'}{2\pi i} \oint\frac{dt}{2\pi i}(t-t_L)(t'-t'_{M})\sqrt{\frac{\d t_L}{\d t}\frac{\d t'_{M}}{\d t'}}
\frac{t^{\half(1-n)}t_L^{\half(1-n)}{t'}^{\half(n-1)}{t'}_{M}^{\half(n-1)}}
{(t-t')(t_L-t')
(t-t'_{M})(t_L-t'_{M})}\times\nonumber\\
&&\times \left[\sqrt{\frac{t-1}{t-x}\frac{t_L-1}{t_L-x}\frac{t'-x}{t'-1}\frac{t'_{M}-x}{t'_{M}-1}}+
\sqrt{\frac{t-x}{t-1}\frac{t_L-x}{t_L-1}\frac{t'-1}{t'-x}\frac{t'_{M}-1}{t'_{M}-x}}\right]
\ee and then integrate over $x$.
   Using the relations \eqref{tlrels} the above can be written as
\be\label{inter1}
&&\oint\frac{dt'}{2\pi i} \oint\frac{dt}{2\pi i}(t-t_L)(t'-t'_{M})\,
\frac{t^{-n}{t'}^{n-1}\,t_L}
{(t-t')(t_L-t')
(t-t'_{M})(t_L-t'_{M})}\times\nonumber\\
&&\times \frac{t_L-t_0}{t-t_0}\frac{t'_{M}-t_\infty}{t'-t_\infty}\left[\frac{t-1}{t_L-x}\frac{t'-x}{t'_{M}-1}+
\frac{t-x}{t_L-1}\frac{t'-1}{t'_{M}-x}\right].
\ee  Note that the integrand is not a meromorphic function as $t_L(t)$ and $t'_{M}(t')$ have branch cuts.
The $t$ contour is around zero and there are no branch cuts in the vicinity of the origin. The $t'$ contour is around infinity. Thus, deforming this integral towards the origin we will encounter the branch cuts. This contour can be split into an integral around the branch cuts and an integral around the origin. Let us analyze the latter part first.

Note that for the $t'$ integral to have a simple pole at $t'=0$ we have to expand the denominator
in the first line of \eqref{inter1} at least to order $n-3$ in $t$. The reason is as follows. Taking $t\to0$ the integral
scales as ${t' }^{n-4}$. Thus, we at least have to get the above mentioned negative power of $t'$ as the other terms in the integrand give
positive powers of $t'$. However, expanding this denominator to order $n-3$ also gives us a simple pole in $t$. Thus, all the other
terms have to be expanded to zeroth order. This in particular implies that the residue will be independent of $x$. Thus, the integral over the moduli space $x$ will vanish as it does for the chiral operators. We deduce that the only non zero contribution to
the one loop comes from $t'$ integrals around the cuts.

There are three cuts for any given $M$: one cut of order $n$ running between $t'=t_0$ and $t'=t_\infty$ and two cuts
of order two running between points satisfying $t'_M(t')=1,\, x$. As we do not know explicitly the functions $t_L$ and $t'_M$
the evaluation of these contour integrals is a complicated task. In principle, one can perform
a consistent expansion in $1/n$ of these functions and try to evaluate the integrals. We leave the detailed investigation
of these issues for future results. However, just from the generic structure
of $t_L(t)$ discussed in section~\ref{tlsec} we can learn that the two point function has the following structure.
\be
\langle M|L\rangle_{1-loop}  =&&f_a(L|M)-f_a(L|M+1)+f_b(L|M)-f_b(L|M-1),\\
&&+f_a(L|-M)-f_a(L|-M)+f_b(L|-M)-f_b(L|-M-1),\nonumber
\ee  where $a,b$ label the two different maps. The first line above comes from the two  order two cuts and
the second line from the order $n$ cut. We have assumed the large $n$ limit and thus that the contour integrals around the two
order two cuts are equal up to a sign. Of course we can also write another expression by exchanging $M$ and $L$.
This expression has a structure of ``nearest neighbor interactions''.  For this to hold precisely 
the functions $f_{a/b}(L|M)$ have to be proportional to $\delta_{L,M}$. To determine whether this is true an explicit
 computation has to be performed. However, in general we have mixing already at tree level and thus we do not have a reason to expect
the above precise property to hold.
This observation can be generalized to any type of impurities. Adding more interaction vertices we will have more cuts
and thus smearing of this nearest neighbor feature.

\section{Summary and Discussion}\label{sumsec}

We have investigated a spin chain picture for the
single-cycle gauge invariant states of ${\rm Sym}^NT^4$. The ground state
of a chain of length $n$ is given by the chiral state $O^{(0,0)}_n$. This state is a bare twist field permuting $n$
copies of $T^4$ appropriately dressed with fermions to render the conformal dimension equal to the $R$-charge.
One considers the different copies (``colors'') the bare twist field permutes as the sites of the spin chain.
In the vacuum state all the ``colors'' entering the bare twist field have the same dressing.
Note that the state with lowest conformal dimension is the bare twist field and not the chiral state one builds from it and which
we take as the vacuum of the spin chain. This is just the
first of many qualitative differences between the symmetric product orbifolds and the gauge theories.
In some sense the ground state is a ``Dirac sea'' of fermions~\cite{Lunin:2002fw} and the impurities are the excitations of this sea.

A natural set of impurities is introduced by changing the dressing of single colors permuted by the twist field. The fields with the lowest
conformal dimension in the theory are the fermions and the basic impurities are given by dressing a site of the spin chain with these.
We have explicitly shown that states with the same quantum numbers but with different impurities mix
already at tree level. The technical reason for this is that on the covering surface the different copies permuted by the twist
fields are identified and there is no suppression of contractions of fields sitting at different sites of the chain. This is in sharp
contrast with  the $1/N$ suppression of such contractions in a gauge theory.
Of course, one can always diagonalize the states at tree level. 
An important open problem is to find whether
 there is a natural and economic way to describe the orthogonal basis of impurities.
 Ultimately one would like to find an operational definition of asymptotic states.

The generators of the superconformal algebra  have an explicit realization
in terms of the basic fields of the theory. Some of these generators, {\it e.g.} the $R$-charge,
are quadratic combinations of the fields. Thus, again unlike the gauge theory case, the
symmetry generators are not the most fundamental impurities. Excitations of the spin chain generated by
the symmetry algebra can be regarded as ``composites'' of more fundamental  impurities.

We have also  discussed higher ``loop'' computations. These are given by turning
on an appropriate twist two interaction and expanding in the coupling constant. We have discussed the first non trivial, ``one loop'',
order in this expansion. The computation is also restricted to the leading order in $1/N$, {\it i.e.} the covering surface is a sphere.
 The explicit evaluation of the one loop reduces to a free field
computation on a covering surface. The fact that the theory is interacting is manifested in two ways. First,
the twist two field insertions give rise to a non trivial covering map. Moreover, the definition of the impurities in terms
of contour integrals when lifted to the covering surface involves a function, $t_L(t)$, which is sensitive to the presence of
the interaction terms. We have shown that the one loop computation of a two point function amounts to evaluating
certain contour integrals around branch cuts of $t_L(t)$. This system of branch cuts has in some  sense a ``nearest neighbor'' structure.
This might be an indication of the local (nearest neighbor) nature of the one-loop Hamiltonian acting
on the appropriate basis of  (tree level) orthogonal   impurities.

The main difficulty of performing explicit higher loop computations in gauge theories is the fast growth in the number of Feynman diagrams.
 In symmetric product CFTs the different ``diagrams'' are the different maps to the covering surface.
As we have shown in section~\ref{mapsec} \textit{all}  the maps relevant  to our problem can be explicitly computed at least
in the limit of large size of the spin chain. It would appear that
 higher loop calculations are not much more complex than one loop calculations, at least for chains of large size.
  So we may hope that a thorough understanding of  one loop  would lead to an  all loop result.

\vskip 1.5 cm

\section*{Acknowledgements}
We thank Antal Jevicki, Samir Mathur, and Luca Mazzucato for useful discussions.
The work of AP was supported in part by DOE grant DE-FG02-91ER40688 and
NSF grant PHY-0643150. The work of LR and SSR is supported in part by a Junior Investigator Award of the Department of Energy
and by a Grant of the National Science Foundation. Any opinions, findings, and conclusions or recommendations expressed in this material are those of the authors and do not necessarily reflect  the views of the National Science Foundation.

\appendix

\section{A short explanation of the diagrams}\label{diagssec}

In this appendix we give a lightening summary of the diagrammatic technique for symmetric product orbifolds introduced in \cite{Pakman:2009zz}.
One way to compute a correlator of twist fields is to lift the computation to the covering surface where the fields are single
valued. There are however different ways to lift a given set of twist fields, {\it i.e.} given set of branching points. The different
maps correspond to different ways to choose common indices, {\it i.e.} common ``colors'', for the twist fields. To obtain a ``gauge''
invariant result one has to sum over all such maps, {\it i.e.} inequivalent choices of the color assignments to the twist fields.
One can think of this sum as a formal sum over diagrams, much like the correlators in a gauge theory are sums over Feynman diagrams.

\begin{figure}[htbp]
\begin{center}
$\begin{array}{c}
\epsfig{file=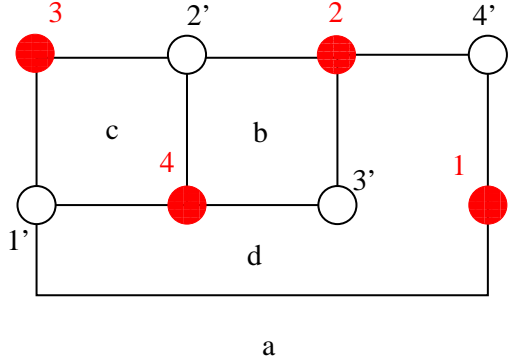,scale=0.4}  \\
\end{array}$
\end{center}
  \caption{A diagram contributing to the correlator $\langle O_{[3]}(z_a)O_{[2]}(z_b)O_{[2]}(z_c)O_{[3]}(z_d)\rangle$
with $z_a<z_b<z_c<z_d$.
The diagram corresponds to  the following choice of (color) indices $(132)_a (24)_b (34)_c (241)_d$.
  The (solid) red dots are the color vertices and the white dots are the non color ones.
  Each {\it loop} (letter) corresponds to a twist field: going around
  the loop counterclockwise one reads off the color indices of
  the corresponding cyclic permutation.} \label{okexample}
\end{figure}

For simplicity we take all the twist fields to correspond to single-cycles.
There are two ways to define the diagrams which are graph theoretic dual of each other. In the body of the paper we use diagrams
in which each twist field corresponds to a loop.  A cycle of length $n$ corresponds to a loop connecting $2n$ vertices. See example
in figure \ref{okexample}. There
are two types of vertices, color and non-color ones, and their position is alternated along each loop. One can assign numbers
to the color vertices and then the cycle-structure can be read off each loop by reading counter clockwise these numbers.
The diagrams are obtained by gluing the loops in all possible ways modulo two rules. The first rule is that in each diagram the number of
color vertices should equal the number of non color ones. The second rule is defined as follows. Each vertex defines a partial
cyclic ordering on the loops, the color vertices by going around them counter clockwise and the non-color ones by going around them
clockwise. The second rule is that all these orderings should be compatible with each other and also compatible with the radial ordering of the
positions of the twist fields. The claim is that there is one to one correspondence between diagrams satisfying these conditions
and the different maps contributing to a correlator of twist fields.

We can think of the loop corresponding to a single-cycle as a spin chain, with the color vertices being
the sites of the chain. The case with two twist-two interaction terms all the possible diagrams are depicted in figures
\ref{diags1} and \ref{diags2}. It is easy to check explicitly that the two basic rules are satisfied.

\section{Diagonalization of the mixing of two $J^-$ impurity states}\label{diagapp}
\begin{figure}[htbp]
\begin{center}
$\begin{array}{c@{\hspace{0.05in}}c}
\epsfig{file=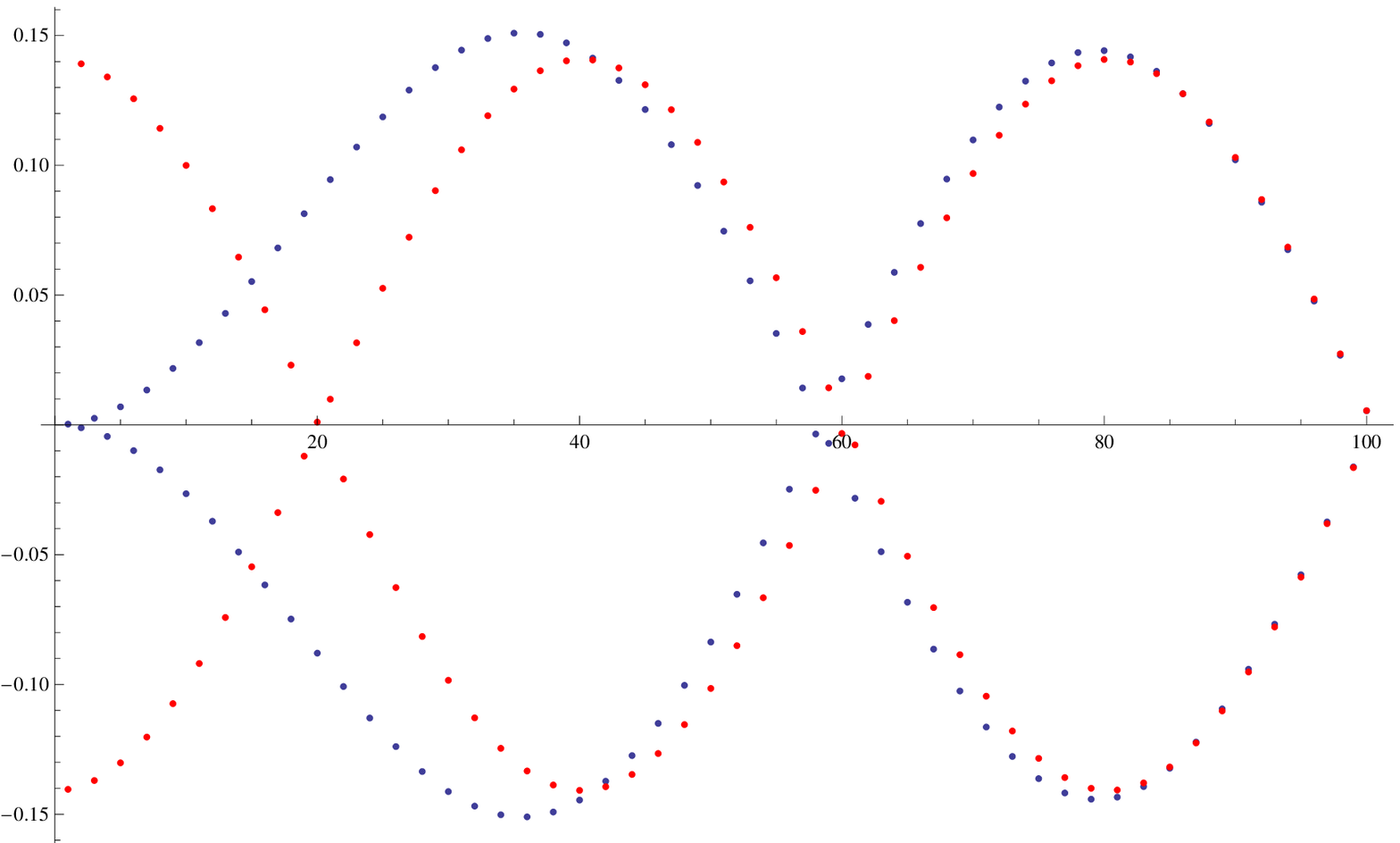,scale=0.4}& \epsfig{file=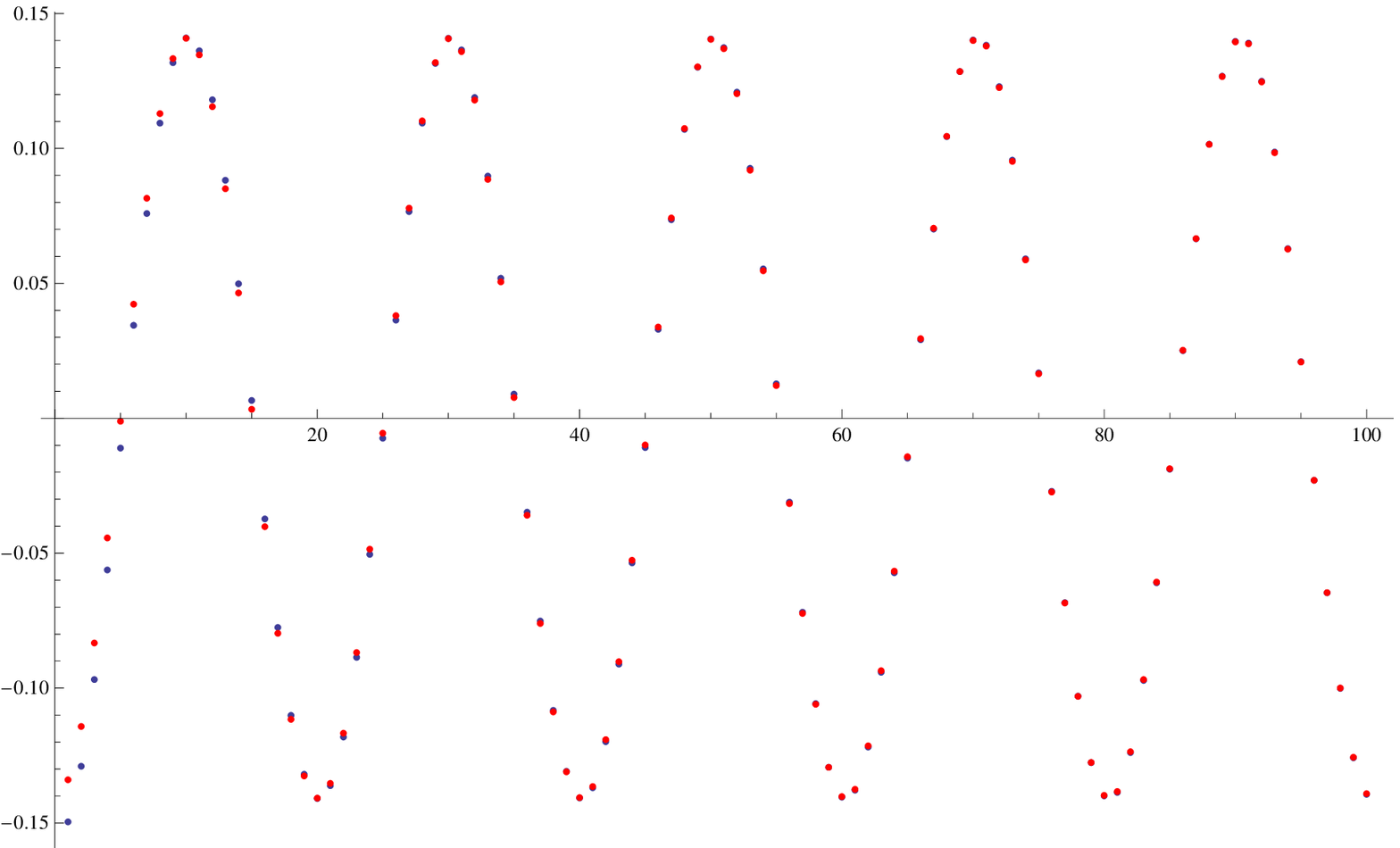,scale=0.4}
\\ [0.2cm]
\end{array}$
\end{center}
 \begin{center}
\caption{The expansion coefficients for two diagonal states, {\it i.e.} on the $x$ axis are the $L$
states and the $y$ axis is the coefficient of each state in the ``momentum'' state decomposition.
 The two states are obtained by taking $k=98$ (left) and
$k=10$ (right) with $n=201$. The blue dots are the expansion coefficients and the red dots are the coefficients only of $U$.
Note that for small $k$ these are approximately identical.
 } \label{diagonalization}
\end{center}
\end{figure}

In this appendix we describe the diagonalization procedure for the tree level mixing matrix \eqref{diagmatrix1}.
 The diagonalization matrix can be conveniently written as a product of three matrices $D\,V\, U$.
The non unitary matrix $U$ is the following Fourier-like transform
\be
{U^k}_{L}=\cos\frac{2\pi k\, L}{n},
\ee where $k$ labels the new states and we can think of it as a ``momentum'' variable. The unitary matrix $V$ is given by
\be
V=
\left(\begin{tabular}{cccccc}
 $\left(\lfloor\frac{n}{2}\rfloor\right)^{-\half}$&$\left(\lfloor\frac{n}{2}\rfloor\right)^{-\half}$&$\left(\lfloor\frac{n}{2}\rfloor\right)^{-\half}$&
$\left(\lfloor\frac{n}{2}\rfloor\right)^{-\half}$&\dots&$\left(\lfloor\frac{n}{2}\rfloor\right)^{-\half}$\\
$0$&\dots&$0$&$0$&$\a_2$&$-\,\a_2$\\
$0$&\dots&$0$&$2\a_3$&$-\a_3$&$-\,\a_3$\\
$0$&\dots&$3\a_4$&$-\a_4$&$-\a_4$&$-\,\a_4$\\
&&\dots&&&\\
$\lfloor\frac{n}{2}\rfloor\a_{\lfloor\frac{n}{2}\rfloor}$&\dots&$-\a_{\lfloor\frac{n}{2}\rfloor}$&$-\a_{\lfloor\frac{n}{2}\rfloor}$
&$-\a_{\lfloor\frac{n}{2}\rfloor}$&$-\a_{\lfloor\frac{n}{2}\rfloor}$\\
\end{tabular}
\right),\qquad \a_k=\frac{1}{\sqrt{k(k-1)}}.\nonumber\\
\ee
 The matrices are $\lfloor\frac{n}{2}\rfloor\times\lfloor\frac{n}{2}\rfloor$ dimensional.
After diagonalizing we also have to rescale
the states with the inverse of the eigenvalues. The eigenvalues of the state $k$ is
\be
k\neq 1\,:\,\e_k=\frac{6\, n^2-16\, n+6}{n^2}-\frac{8}{n^2}\,(k-1)\,k\, ,\qquad k=1 \,:\, \e_1=4(n-2).
\ee That is we define a matrix $D$ by
\be
D_{k\, k'}=\sqrt{\e_k}\,\delta_{k k'} .
\ee
The matrix ${\mathcal I}$ transforms as
$\left(D^{-1}\,V\, \left(U^\dagger\right)^{-1}\right)\;{\mathcal I}\;\left(U^{-1}\,V^\dagger\, D^{-1}\right)=Id$.
Note that after acting only with $U$ the small momentum states are approximately orthogonal in the large $n$ limit,
a fact illustrated in  figure \ref{diagonalization}.

\section{Some details of the one loop map}\label{detailssec}
In this appendix we give some formulae needed for the one loop computation in section \ref{vacsec}.
Lets us find the expansion of $z-u$ in terms of
$y=t-x$, {\it i.e.} the inverse map. We can write the expansion as
\be y=\sum_{k=1}^\infty c_k(z-u)^{k/2}.
\ee
To obtain the coefficients of the inverse map we first write the following
\be
\ln \frac{z(t)}{u} =n\ln \frac{t}{x}+\ln \frac{t-t_0}{x-t_0}-\ln \frac{t-t_\infty}{x-t_\infty},
\ee and expand both sides to get
\be
\sum_{k=1}^\infty\frac{(-1)^{k+1}}{k}\left(\frac{z-u}{u}\right)^k=(t-x)^2\sum_{k=0}^\infty  a_k\,(t-x)^{k},
\ee with
\be
a_k=\frac{(-1)^{k+1}}{k+2}\left(\frac{n}{x^{k+2}}+\frac{1}{(x-t_0)^{k+2}}-\frac{1}{(x-t_\infty)^{k+2}}\right).
\ee
The expansion coefficients are related as
\be
c_1^2=\frac{1}{u(x)\,a_0},\qquad c_2=-\frac{a_1}{2u(x)\,a_0^{2}},\qquad
c_3=-\frac{2a_0^3-5a_1^2+4a_0a_2}{8u(x)^2a_0^4\,c_1}.
\ee

The above results are needed to write down the differential equation~\eqref{diffeq}.
We get for the quantities appearing in this equation,
\be
\left(\frac{t''}{t'}\right)'&=&\frac{1}{2}\frac{1}{(z-u)^2}-
\frac{c_2}{c_1}\frac{1}{2}(z-u)^{-3/2}+\dots\\
\left(\frac{t''}{t'}\right)^2&=&\frac{1}{4}\frac{1}{(z-u)^2}-
\frac{c_2}{c_1}(z-u)^{-3/2}+3
\left[\frac{c_2^2}{c_1^2}-\frac{c_3}{c_1}\right](z-u)^{-1}+\dots
\nonumber
\ee From here we obtain
\be
\left(\frac{t''}{t'}\right)'-\half \left(\frac{t''}{t'}\right)^2&=&
\frac{3}{8}\frac{1}{(z-u)^2}-
\frac{3}{2}\left[\frac{c_2^2}{c_1^2}-\frac{c_3}{c_1}\right](z-u)^{-1}+\dots .
\ee We will also need the following
\be
(t')^2&=&
\frac{c_1^2}{4}(z-u)^{-1}+c_1c_2(z-u)^{-1/2}+\frac{1}{2}\left(2c_2^2
+3c_1c_3\right)\dots
\nonumber\\
\frac{1}{t-x}&=&
\frac{1}{c_1}(z-u)^{-1/2}-\frac{c_2}{c_1^2}+\frac{c_2^2-c_1c_3}{c_1^3}(z-u)^{1/2}+\dots\\
\frac{1}{t-x+a}&=&
\frac{1}{a}-\frac{c_1}{a^2}(z-u)^{1/2}-\left[\frac{c_2}{a^2}-\frac{c_1^2}{a^3}\right](z-u)+
\frac{-c_1^3 + 2 a c_1 c_2 - a^2 c_3}{a^4}(z-u)^{3/2}+\dots\;,\nonumber
\ee where $a$ is some complex number.

\section{Arutyunov-Frolov map}\label{AFMap}
Note that the one loop map of section \ref{mapsec}, although useful to understand the structure of different diagrams, is complicated as
it contains square roots. Essentially,  the map can be recast in much simpler
form, as  has been done by Arutyunov and Frolov in~\cite{Arutyunov:1997gt}. The simplification occurs if
we map the twist at $z=\infty$ to $t=\infty$ and the twist $z=0$ to $t=0$ as before, but
we map the additional image of $z=0$ to $t=x-1$ (instead of mapping the twist at $z=1$ to $t=1$ as was done in section \ref{mapsec}). Then the map is given by the following
\be
z(t)=t^n\;\frac{t-t_0}{t-t_\infty}\;\frac{t_1-t_\infty}{t_1^n\,(t_1-t_0)},
\ee  where we have
\be\label{ts}
&&t_0=x-1,\qquad t_\infty=x-\frac{x}{x+n},\\
&&t_1=\frac{1-n}{n}+x-\frac{n+1}{n}\;\frac{x}{x+n}.
\ee With these definition the point $z=u$ maps to $t=x$ with the following relation between the two
\be\label{modAF}
u=v(x)=\frac{x^{n-1}(x+n)^{n+1}}{(x-1)^{n+1}(x+n-1)^{n-1}} \, .
\ee
Note, in contrast to~\eqref{uofx}, that there are no square roots in this expression. Let us understand the moduli space in the new
coordinates. The images of different OPE limits are
\be\label{AFlimits}
u\to 0&\qquad :\qquad & u\sim \frac{n^{n+1}}{(n-1)^{n-1}}x^{n-1} ,\qquad u\sim \frac{n^{n-1}}{(-n-1)^{n+1}}(x+n)^{n+1},\\
u\to \infty&\qquad :\qquad & u\sim \frac{(1+n)^{n+1}}{(n)^{n-1}}\frac{1}{(x-1)^{n+1}} ,
\qquad u\sim \frac{(n-1)^{n-1}}{n^{n+1}}\frac{1}{(x+n-1)^{n-1}},\nonumber\\
u\to 1&\qquad :\qquad & u-1\sim-\frac{64}{3}\frac{n}{(n^2-1)^2}(x-\frac{1-n}{2})^3,\qquad x\to \infty \nonumber
\ee In the limit $u\to 1$ we wrote down only those solutions contributing
to the OPE limit, {\it i.e.} also $t_1\to x$.

We can calculate the first correction to the two point functions of the chiral states using this map to obtain
\be\label{G0AF}
G_0(x)= C\;x^{2-n}\,\left(x-1\right)^{3 + n}\,\,\left(n+x\right)^{1-n}\,\left(x+n-1\right)^{n}\,
\left(x+\frac{n-1}{2}\right)^{-4},
\ee where $C$ is an overall constant not fixed by our method
of computation. A simple check of this equation is to take $x\to\infty$. In this limit $u\to1$ and
corresponds to an OPE limit of the two interactions in the way that they share all the colors
and thus annihilate each other, see equation \eqref{AFlimits}.
 The four point function scales as $G(u)\sim \frac{1}{(1-u)^2}$ as expected as the interactions have dimension one. Computing the coefficient of the
 subleading term in the $OPE$, {\it i.e.} the term coming from dimension one operator, we find that it is zero. Thus there is no contribution from possible
contact terms from the untwisted sector.
In another limit, $x\to \frac{1-n}{2}$, we have $u\to 1$ and the leading singularity is $-4$. This gives the conformal dimension
of the operator of the leading $OPE$ singularity to be $\Delta=\frac{2}{3}$. This is the dimension of bare twist three field.
Farther, expanding to subleading order, {\it i.e.} $(x+\frac{n-1}{2})^{-3}$, we get that the coefficient is zero. This subleading order
corresponds to $\Delta=1$. Thus, the vanishing of this coefficient implies either that the correlator of the chiral states
with the contact, twist three, terms vanishes, or that the contact terms simply do not exist.

We can easily integrate the above expression over the location of the interaction
\be
\sum_{j=1}^{2n}\int d^2 u G(x_j(u),\bar x_j(\bar u))&=&\int d^2x |v'(x)|^2 G(x)G(\bar x)=
\int d^2x \left|v'(x) G(x)\right|^2=\\
&=&|nC|^2\int d^2x \left|\frac{(x-1) (n + x)}{ (n-1 + 2 x)^2}\right|^2\nonumber
\ee We now make the following change of variables
\be
y=-4\frac{(x-1) (n + x)}{(1 + n)^2},
\ee we obtain
\be \label{finintAF}
|nC|^2\int d^2x \left|\frac{(x-1) (n + x)}{ (n-1 + 2 x)^2}\right|^2=
\left(\frac{n^2 (1 + n)|C|}{16}\right)^2\int d^2y\frac{ |y|^2}{|1 - y|^3}
\ee Using the following
\be
\int d^2z\;z^a{\bar z }^{\bar a}(1-z)^b(1-\bar z)^{\bar b}=\pi\frac{\Gamma(1+a)\Gamma(1+b)\Gamma(-\bar a-\bar b-1)}
{\Gamma(-\bar a)\Gamma(-\bar b)\Gamma(a+b+2)},
\ee and plugging $a=\bar a=1$, $b=\bar b=-\frac{3}{2}$ we get that \eqref{finint} is vanishing
and thus the chiral operators do not acquire an anomalous dimension.

\section{The  one loop correlator of bare twist fields}\label{bosmap}
Using map $a$ and computing the four point function of twist fields without the dressings we obtain
\be
\ln G(u(x))=\frac{1}{8n}\biggl(&&-2 n \ln( x-1) - (2 + 5 ( n-1) n) \ln x -
   2 n \ln(1 + x + \sqrt{n^2 ( x-1)^2 + 4 x}) +\nonumber\\&&+ (2 +
      n ( 2 n-3)) \ln(
     n^2 ( x-1) - 2 x - n \sqrt{n^2 ( x-1)^2 + 4 x}) +\nonumber\\&&+ (2 +
      n (3 + 2 n)) \ln(
     2 + n (n ( x-1) + \sqrt{n^2 (x-1)^2 + 4 x}))\biggr).
\ee Remembering that the conformal dimension of a twist field
is
\be
\Delta_n=\frac{1}{4}\left(n-\frac{1}{n}\right),
\ee we can check different OPE limits. For instance as $u\to 1$ and $x\to 1$ we get that
\be
\ln G(u)\sim -\frac{1}{4}\ln(x-1)\sim-\frac{1}{12}\ln (u-1),\ee and we get also
\be
\frac{1}{12}=\Delta_2+\Delta_2-\Delta_3.
\ee Taking the limit $u\to 0$ and $x\to 0$ we get
\be
\ln G(u)\sim -\frac{2 + 5 ( n-1) n}{8n}\ln x\sim -\frac{2 + 5 ( n-1) n}{8n(n-1)}\ln u,\ee which is to be understood as
\be
 \frac{2 + 5 ( n-1) n}{8n(n-1)}=\Delta_n+\Delta_2-\Delta_{n-1}.
\ee
In the same way for map $b$ we obtain
\be
\ln G(u(x))=-\frac{1}{8n}\biggl(&&6 n \ln( x-1) + ( n + n^2-2) \ln x -
   2 n \ln(1 + x + \sqrt{n^2 ( x-1)^2 + 4 x}) +\nonumber\\&&+ (2 +
      n ( 2 n-3)) \ln(
     n^2 ( x-1) - 2 x - n \sqrt{n^2 ( x-1)^2 + 4 x}) +\nonumber\\&&+ (2 +
      n (3 + 2 n)) \ln(
     2 + n (n ( x-1) + \sqrt{n^2 (x-1)^2 + 4 x}))\biggr).
\ee The  different OPE limits give here the following. As $u\to 1$ and $x\to 1$ we get that
\be
\ln G(u)\sim -\frac{3}{4}\ln(x-1)\sim-\frac{3}{4}\ln (u-1),\ee which is consistent with
\be
\frac{3}{4}=\Delta_2+\Delta_2,
\ee the two interactions annihilate each other as they share both their colors. Taking the limit $u\to 0$ and $x\to 0$ we get
\be
\ln G(u)\sim -\frac{n^2+n-2}{8n}\ln x\sim -\frac{n^2+n-2}{8n(n+1)}\ln u,\ee and we get also
\be
 \frac{n^2+n-2}{8n(n+1)}=\Delta_n+\Delta_2-\Delta_{n+1}.
\ee Thus this computation gives us the identification between the two maps and the two copies of the moduli space.

\bibliography{h3bib}{}
\bibliographystyle{JHEP}

\end{document}